\documentclass{aa}

\usepackage{natbib}
\usepackage{caption}
\usepackage{subcaption}
\usepackage[flushleft]{threeparttable}
\usepackage{graphicx}
\usepackage{txfonts}
\usepackage{xcolor}
\usepackage{euscript}
\usepackage{lscape}

\usepackage{hyperref}

\begin{document} 

\defcitealias{piras2024a}{Paper~I}  

\title{LOFAR Deep Fields: Probing the sub-mJy regime of polarized extragalactic sources in ELAIS-N1}
\subtitle{II. Analysis}

   \author{S. Piras\inst{\ref{inst0}}
        \and C. Horellou\inst{\ref{inst1}}
        \and J.E. Conway\inst{\ref{inst1}}
        \and M. Thomasson\inst{\ref{inst0}}
        \and T.W. Shimwell\inst{\ref{inst2},\ref{inst3}}
        \and S.P. O'Sullivan\inst{\ref{inst4}} 
        \and E. Carretti\inst{\ref{inst5}}
        \and V. Vacca\inst{\ref{inst7}}
        \and A. Bonafede\inst{\ref{inst5}, \ref{inst8},}
        \and I. Prandoni\inst{\ref{inst5}}
          }
    \institute{Department of Space, Earth and Environment, Chalmers University of Technology, 412 96 Gothenburg, Sweden\label{inst0}
   \and Department of Space, Earth and Environment, Chalmers University of Technology, Onsala Space Observatory, 43992 Onsala, Sweden\label{inst1} 
   \and ASTRON, Netherlands Institute for Radio Astronomy, Oude Hoogeveensedijk 4, 7991 PD, Dwingeloo, The Netherlands \label{inst2} 
   \and Leiden Observatory, Leiden University, P.O. Box 9513, 2300 RA Leiden, The Netherlands  \label{inst3} 
    \and Departamento de Física de la Tierra y Astrofísica \& IPARCOS-UCM, Universidad Complutense de Madrid, 28040 Madrid, Spain\label{inst4}
  \and INAF Istituto di Radioastronomia, Via Gobetti 101, I-40129 Bologna, Italy \label{inst5} 
 \and INAF-Osservatorio Astronomico di Cagliari, Via della Scienza 5, I-09047 Selargius (CA), Italy \label{inst7} 
 \and DIFA Universitá di Bologna, via Gobetti 93/2, 40129 Bologna, Italy \label{inst8}
             }

  \date{Received ...} 
 
 
  \abstract
   {
   Deep polarization surveys at low radio frequencies are key to cosmic magnetism studies: Larger catalogs of polarized extragalactic sources and increased precision on Faraday rotation measures (RMs) make it possible to probe the magneto-ionic medium along the lines of sight of the sources and to construct denser RM grids. 
   In a first paper, we presented a search for polarized sources in deep observations of the 25-square-degree area of the 
   European Large Area ISO Survey-North 1 (ELAIS-N1) field with the LOw Frequency ARray (LOFAR) at 114.9–177.4 MHz. 
   }
   {In this paper, we investigate the properties of the polarized radio galaxies and use the catalog to produce an RM grid of the field.}
   {After identifying the host galaxies and collecting redshift information, we characterized the radio galaxies in terms of their radio morphologies, rest-frame radio luminosities, and linear sizes. We calculated residual rotation measures (RRMs) by removing the Galactic RM and studied the variation in the RRMs with redshift and degree of polarization. We produced an RRM grid of the field and compared the positions of the polarized sources with those of galaxy clusters and superclusters.}
   {The radio galaxies show a variety of morphologies, including diffuse emission; Fanaroff Riley type II sources make up about half of the sample. Using available multiband catalogs, we found redshifts for the hosts of all polarized sources in the range of 0.06 to 1.9. Polarized emission is detected mainly from large radio galaxies. 
   The RRM values have a median close to zero, and they appear to be independent of redshift and degree of polarization. 
   The sources in the lines of sight of clusters of galaxies and of a supercluster are indistinguishable in their polarization and RRM properties from the population of sources that are not behind these structures.
   }
   {}


   \maketitle

\section{Introduction}
\label{sec:int}

\begin{table*}[t]
\caption{Properties of the LoTSS-DR2 RM catalog and of the ELAIS-N1 LOFAR Deep Field catalog.}
\centering                         
\label{table:cats}      
\begin{tabular}{l c c c c c c c c c c}        
\hline\hline  
 Catalog & Reference & Area     & Resolution & $\nu_{\rm c}$ & $\Delta\nu$ &$\delta\phi$ 
         & $\sigma_{\rm QU}$   & $N_{\rm RM}$ & $N_{\rm RG}$ & $N_{\rm z}$\\ 
         &          & (deg$^2$) & (arcsec) &   (MHz)    & (MHz) & (rad~m$^{-2}$)
 &($\mu$Jy beam$^{-1}$) \\
\hline
LoTSS-DR2 RM       &(1)   &5720 &20 & 144  &48  & 1.16 & 75    &2461 & 1980 & 1783\\
ELAIS-N1           &(2)   & 25  & 6 & 146  &63  & 0.9  & 20    & 33  & 31 & 31\\
\hline                                 
\end{tabular}
\tablefoot{
(1) \cite{lotssdr2rm}. The quoted $\sigma_{\rm QU}$ noise value corresponds to the median detection threshold, 8~$\sigma_{\rm QU} = 0.6$~mJy~beam$^{-1}$. 
(2) \citetalias{piras2024a}. 
The quoted noise value is the value measured at the center of the ELAIS-N1 field. 
$\nu_c$ is the central frequency, $\Delta\nu$ is the bandwidth, $\delta\phi$ is the resolution of the Faraday spectra, and $\sigma_{\rm QU}$ is the noise level in the Stokes $Q$ and $U$ Faraday cubes (\citetalias{piras2024a}). 
$N_{\rm RM}$ is the number of entries in the RM catalogs. 
$N_{\rm RG}$ is the number of radio galaxies associated with at least one RM value. 
$N_{\rm z}$ is the number of radio galaxies with a known redshift and associated with at least one RM value.
}
\end{table*}

Observations of polarized synchrotron radiation at radio frequencies are the most direct probes of cosmic magnetism. As radiation from an extragalactic source traverses a magneto-ionic medium, its plane of polarization experiences Faraday rotation, an effect that is proportional to the square of the wavelength of the EM wave. 
The observed rotation measure (RM; expressed in rad~m$^{-2}$), is related to the 
product of the thermal electron density, $n_{\rm e}$, and the line-of-sight component of the magnetic field, $B_{\parallel}$, integrated along the line of sight, 
\begin{equation}
\left( \frac{\rm RM}{{\rm rad~m}^{-2}} \right) 
= 0.812 \int_{\ell = 0}^{\ell = L} 
\left( \frac{n_{\rm e} (\ell)} { {\rm cm}^{-3} }\right) 
\left( \frac{B_{\parallel}(\ell)} {\mu{\rm G}} \right)
\left( \frac{{\rm d}\ell}{\rm pc}  \right) 
\, .
\label{eqphi}
\end{equation}

The observed RM depends on all the contributing magneto-ionic media between the source and the observer, including the Galactic interstellar medium, the intergalactic medium, intervening galaxies, and the plasma within the source itself. 
So-called RM grids (i.e., collections of RM values across the sky) are a potentially powerful tool for investigating magnetic fields. 
To distinguish the different contributions to the RMs, 
it is crucial to identify the optical and/or IR counterparts to the polarized radio sources, however, and to obtain redshift information on the host galaxies 
to characterize the sources that constitute the RM grid in terms of their radio morphologies, their sizes, and other properties.

Traditionally, RM grid studies were conducted at 1.4~GHz and were hampered by large systematic uncertainties because the frequency sampling was poor. The largest RM catalog available to date is that of \cite{taylor2009}, which was produced from the National Radio Astronomy Observatory (NRAO) Very Large Array (VLA) Sky Survey  (NVSS; \citealt{condon1998}). The catalog contains RM values for $37\,543$ lines of sight to polarized radio sources and covers the sky north of a declination $-40^\mathrm{o}$ at a resolution of 45~arcsec (82\% of the whole sky). Through cross-matching with redshift catalogs, \cite{Hammond2012arXiv1209.1438H} identified a total of 4003 matches for the NVSS RM catalog.
More recently, a new generation of telescopes operating at lower frequencies has been used to  map the sky and obtain more precise RM values through low-frequency broadband polarimetry. In particular, at the LOFAR Two-Metre Sky Survey (LoTSS; \citealt{Shimwell2017, Shimwell2019}) provided data at 144 MHz with a resolution of 20~arcsec that were used to produce the LoTSS-DR2 RM catalog \citep{lotssdr2rm}, a collection of $\sim$2500 high-precision ($\sim 1$~rad~m$^{-2}$) RM values from extragalactic polarized sources over 5720~deg$^2$ of the northern sky (Table~\ref{table:cats}). The catalog also contains host galaxy identifications for 88\% of the sources, along with redshifts for 79\%. 

Observations at different wavelengths and of sources at different redshifts are required to distinguish the different RM contributions. 
\cite{Vernstrom2019ApJ...878...92V} studied differences in RMs between adjacent NVSS sources on the sky (1.4~GHz; \citealt{condon1998, taylor2009}) and concluded that these variations may indicate the detection of an extragalactic RM signal.
\cite{OSullivan2020} applied the method of \cite{Vernstrom2019ApJ...878...92V} to the LoTSS (144~MHz, \citealt{Shimwell2019}) and placed an
upper limit of 4~nG on the cosmological comoving magnetic field strength on megaparsec (Mpc) scales. \cite{Carretti2022} compared data from the NVSS and LoTSS-DR2 RM catalogs and found that the observed residual RM (RRM; obtained after subtraction of the Galactic RM) is most likely to have an origin local to the source at 1.4~GHz, while a cosmic web filament origin is favored at 144~MHz. This shows that RM studies at low radio frequencies have important implications on our understanding of the processes related to magnetic fields in the Universe \citep{Carretti2023}.

Depolarization is the primary challenge in low-frequency polarization studies (e.g., \citealt{Stuardi2020}, \citealt{Mahatma2021}), hence the need for deep polarization surveys. 
The ELAIS-N1 LOFAR Deep Field \citep{Sabater2021} is best suited for a deep search for polarization. 
\cite{HerreraRuiz2021} stacked LOFAR polarization datasets from observations of this field that were carried out at six epochs. They reached a 1-$\sigma$ sensitivity of 26~$\mu$Jy~beam$^{-1}$ in the final Stokes $Q$ and $U$ images at a resolution of 20$''$ and detected ten polarized sources in an area of 16~deg$^2$.

In \cite{piras2024a} (hereafter \citetalias{piras2024a}), 
we developed a new method to stack LOFAR datasets that were taken in different observing cycles with different frequency configurations, and we applied the method to the ELAIS-N1 LOFAR Deep Field to produce the most sensitive polarization dataset at 150~MHz so far. Stacking polarization datasets from 19 epochs, we reached the lowest noise level of $\sigma_{\rm QU} \sim 20\, \mu$Jy~beam$^{-1}$ in the central region of the final 25~deg$^2$ image of the ELAIS-N1 field at a resolution of 6~arcsec (the deep Stokes $I$ image of \citealt{Sabater2021} was produced by combining data from 22 epochs, or an effective observing time of 163.7 hours, and had a similar noise level). 
We compiled a catalog of 31 polarized radio galaxies, 25 of which were identified through a systematic search across the entire field (setting a threshold for the  signal-to-noise ratio of 8), 
while the remaining sources were identified through a study of the sources known to be polarized at 1.4~GHz from the works of \cite{grant2010} and \cite{taylor2009}. For these sources, we lowered our detection threshold to $6\sigma_{\rm QU}$. We derived the polarized source counts and modeled them based on the total flux density counts obtained by \cite{Mandal2021} in their study of three LOFAR Deep Fields that included ELAIS-N1.
Some properties of the ELAIS-N1 LOFAR Deep Field RM catalog and of the LoTSS-DR2 RM catalog are summarized in Table~\ref{table:cats}. 

The ELAIS-N1 LOFAR Deep Field \citep{Sabater2021}
has rich multiwavelength ancillary data that cover a broad range of the electromagnetic spectrum 
from X-ray to radio bands in its central region (inner 7.15 deg$^2$). 
Through this wealth of information, the host galaxies of radio sources could be  identified \citep{kondapally2021}, and photometric redshifts were derived whenever spectroscopic redshifts were not available \citep{Duncan2021}.
With its exceptional depth and image quality, the ELAIS-N1 LOFAR Deep Field presents an optimal setting for diverse scientific investigations, such as 
the search for diffuse emission within clusters of galaxies \citep{Osinga2021A&A...648A..11O}, the examination of an alignment in radio galaxies \citep{Simonte2023A&A...672A.178S}, 
the search for giant radio galaxies \citep{Simonte2024A&A...686A..21S}, 
and studies of the spectral properties of star-forming galaxies \citep{An2024MNRAS.528.5346A}
and of the Galactic polarized emission \citep{Snidaric2023}. 

In this paper, we exploit the ELAIS-N1 multiband information to characterize the extragalactic polarized sources presented in \citetalias{piras2024a} and compare their properties with those of the LoTSS-DR2 RM catalog (Sect.~\ref{sect:properties}). 
Because LOFAR data lack an absolute polarization calibration, 
we present no relation between the intrinsic (zero-wavelength) polarization angles and the source major-axis position angles, for example; only the rotation of the polarization angles with frequency can be measured.  
In Sect.~\ref{sect:RM_RRMgrid} we present an analysis of the RM and RRM grids of the ELAIS-N1 LOFAR Deep Field, and we conclude in Sect.~\ref{sec:conclusion}. 

In the calculations of the rest-frame luminosities and projected linear sizes of the radio galaxies from which polarization was detected, we used the following values of cosmological parameters for a flat $\Lambda$ cold dark matter model: 
$H_0 = 67.8$~km~s$^{-1}$~Mpc$^{-1}$ and $\Omega_{\rm m} = 0.308$ \citep{PlanckCollaboration2016}.

\section{Properties of the detected polarized sources} \label{sect:properties}

\begin{table*}[t]
\caption{Radio morphological classification and some properties of the polarized radio galaxies in the ELAIS-N1 LOFAR Deep Field.}
\centering                         
\label{table:summaryResults}      
\begin{tabular}{l r l l }        
\hline\hline  
Type    & $N$ & Fraction & ID of polarized component in \citetalias{piras2024a}\\
\hline
Compact & 4      &$12.9\pm6.5$\%     &09, 10, 24, 25\\  
FRI     & 4      &$12.9\pm6.5$\%     &06, 14, 18, 32\\ 
FRII    & 18     &$58.1\pm 13.7$\%   &01, 02, 04$_{\rm A}$ + 04$_{\rm B}$; 05, 07, 08, 11, 12; 13$_{\rm A}^a$ + 29; 15, 17, 19, 21, 23, 26, 27$^b$, 30, 31$^b$\\

Other Extended & 5 &$16.1\pm7.2$\%    &03, 16, 20$^c$, 22, 28\\
\hline
Total   &31     &100\%\\
\hline
Radio size $> 500$ kpc  & 9 & $29.0\pm 9.7$\%     & 08, 11, 12; 13$_{\rm A}$ + 29; 15, 18, 23, 26, 32 \\
Radio size $> 1$ Mpc    &3 & $9.7\pm 5.6$\% & 11, 12, 23\\
\hline
\end{tabular}
\tablefoot{
$N$ is the number of radio galaxies. 
The uncertainties associated with the fractional number of sources of a given type are Poissonian ($\sqrt{N}/31$).\\
a) Sources~13$_{\rm{A}}$ and 29 (13$_{\rm{B}}$) are two lobes of the same FRII radio galaxy.\\
b) Source~27 and 31 were classified as FRIs by \cite{Mingo2022}, but after visual inspection, we classified them as FRIIs.\\
c) Source~20 appears to be a one-sided blazar. 
}
\end{table*}

In this section, we examine some key properties of the radio galaxies in ELAIS-N1 for which polarization was reported in \citetalias{piras2024a}: radio morphology, redshift of their hosts, projected linear size, and rest-frame luminosity. These properties are compared with those of the radio galaxies in the LoTSS-DR2 RM catalog of \cite{lotssdr2rm}. 

\subsection{Radio morphology} 

In Appendix~\ref{app:sources_images} (available on Zenodo\footnote{\url{https://zenodo.org/records/14012234} 
})
we show the $6''$ resolution LOFAR Stokes~$I$ images of the sources in which polarization was detected and the corresponding  Faraday spectra at the pixel of the peak polarized intensity. 
The images were extracted from the large Stokes~$I$ image of \cite{Sabater2021}. 
The pixels in blue are the positions at which polarized intensity was detected in the range indicated by the color scale. 
The polarized emission is unresolved in all the detected sources. 
A description of each source is given in Appendix~\ref{app:sources}. 

The radio sources present a variety of morphologies that we classified into four categories: compact, Fanaroff–Riley types I and II (FRI and FRII; \citealp{FanaroffRiley1974MNRAS.167P..31F}), and other extended sources. 
FRI radio galaxies are center-brightened and become fainter toward the outer extremities of the lobes, 
FRII radio galaxies are edge-brightened and generally terminate in hotspots, 
sources classified as compact 
are those whose full width at half maximum (FWHM) of the deconvolved major axis is smaller than or equal to $6''$ in the \cite{Sabater2021} catalog, 
and the other extended sources are those whose deconvolved FWHM is greater than $6''$, which cannot be classified as FRI or FRII galaxies.
 
In the central region of the ELAIS-N1 LOFAR Deep Field with rich multiwavelength information 
(shown in Fig.~1 of \citetalias{piras2024a}), we detected polarization from 18 radio galaxies; 10 of them are listed as FRI or FRII in the catalog of 
\cite{Mingo2022}\footnote{
\cite{Mingo2022} used the LoMorph Python code {\url{https://github.com/bmingo/LoMorph}} of \cite{Mingo2019} to classify FRI and FRII radio galaxies in the multiwavelength region of ELAIS-N1. 
Ten radio galaxies of the \cite{Mingo2022} sample are associated with our polarized components 08, 11; $13_{\rm A}$ and $13_{\rm B}$; 
18, 19, 20, 26, 27, 29 ($=13_{\rm B}$), 31, and 32.
}. 
We classified the remaining sources in the field through visual inspections of the LOFAR Stokes $I$ image with a resolution of 6 arcsec of \cite{Sabater2021} and of images from the Very Large Array Sky Survey (VLASS). 
We also inspected images of the 10 polarized sources in our sample that were morphologically classified by \cite{Mingo2022}, and we reclassified two of them, as indicated in the footnote of Table~\ref{table:summaryResults}. 

In Table~\ref{table:summaryResults} we present the result of the 
classification of the radio galaxies from which we detected polarization. 
About half of the sources are classified as FRII radio galaxies, and the other half are about equally distributed among the other classes (four to six sources in each category). 

The fraction of polarized radio sources that are classified as FRI galaxies is lower in the ELAIS-N1 LOFAR Deep Field 
($\sim$13\%) than in the LoTSS-DR2 RM sample ($\sim$20\%).
The fraction of detected polarized FRIIs is higher in ELAIS-N1 
($\sim$58\%) than in LoTSS-DR2 RMs ($\sim$40\%). 
However, this difference may not be significant when Poisson\-ian errors are taken into account ($\sim58\pm14$\% in ELAIS-N1) and considering that approximately 15\% of the LoTSS-DR2 RM sources were classified as hybrid when their FR category could not be reliably assessed.

We also examined the fraction of FRI or FRII radio galaxies from which we detected polarization in the multiwavelength part of the field studied by \cite{Mingo2022}, who identified  160 FRI and 126 FRII radio galaxies in total intensity (with angular sizes greater than $27''$ and  at $z < 2.5$). 
As mentioned above, some of the sources may be classified differently after a detailed visual inspection. 
The fact that we detected polarization from more FRII than FRI radio galaxies in the multiwavelength region in which \cite{Mingo2022} found more FRI radio galaxies is an indication that the detection rate of polarization from FRII radio galaxies is higher than that from FRIs.  

In ELAIS-N1, polarization from FRII radio galaxies is detected in lobes or hotspots that are located at large projected distances from their host galaxies, and 
in FRIs, polarization is found in the more central regions, likely from inner jets (this might be a selection effect, however). 
The larger fraction of detected polarized FRIIs might indicate different depolarizing environments through which the polarized emission travels. The lobes of FRIIs may expand in lower-density ionized environments, which are characterized by weaker magnetic fields that cause less depolarization (e.g., \citealt{OSullivan2020, Stuardi2020, Carretti2022}). The central regions of FRIs, in contrast, may be in denser magneto-ionic environments and are therefore more strongly affected by depolarization.

\begin{figure*} [t] 
\centering
\includegraphics[width=0.45\linewidth]{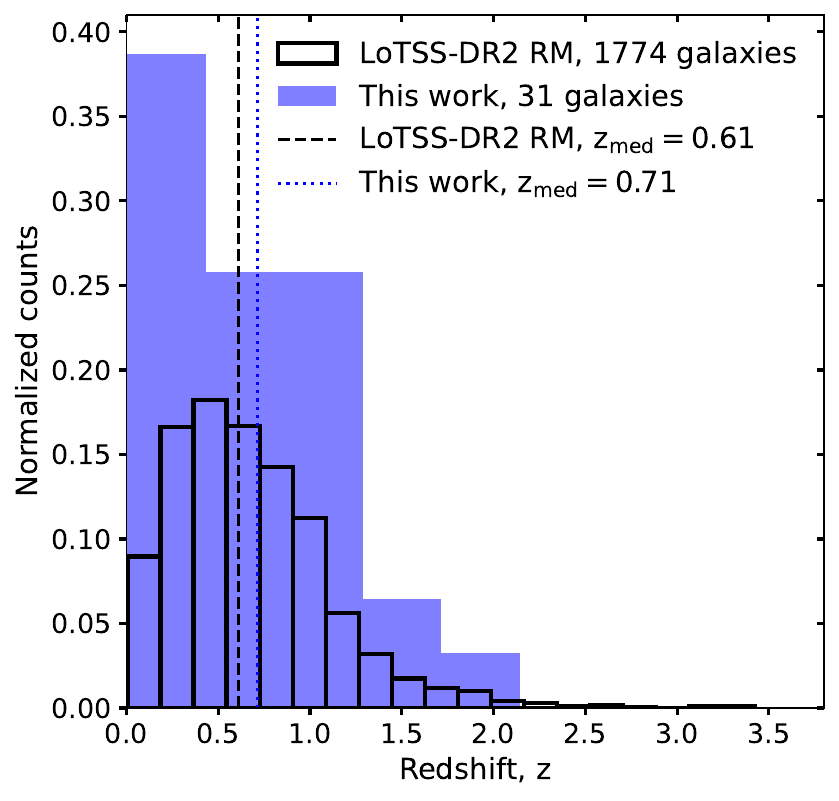} 
\hspace{1cm}
\includegraphics[width=0.45\linewidth]{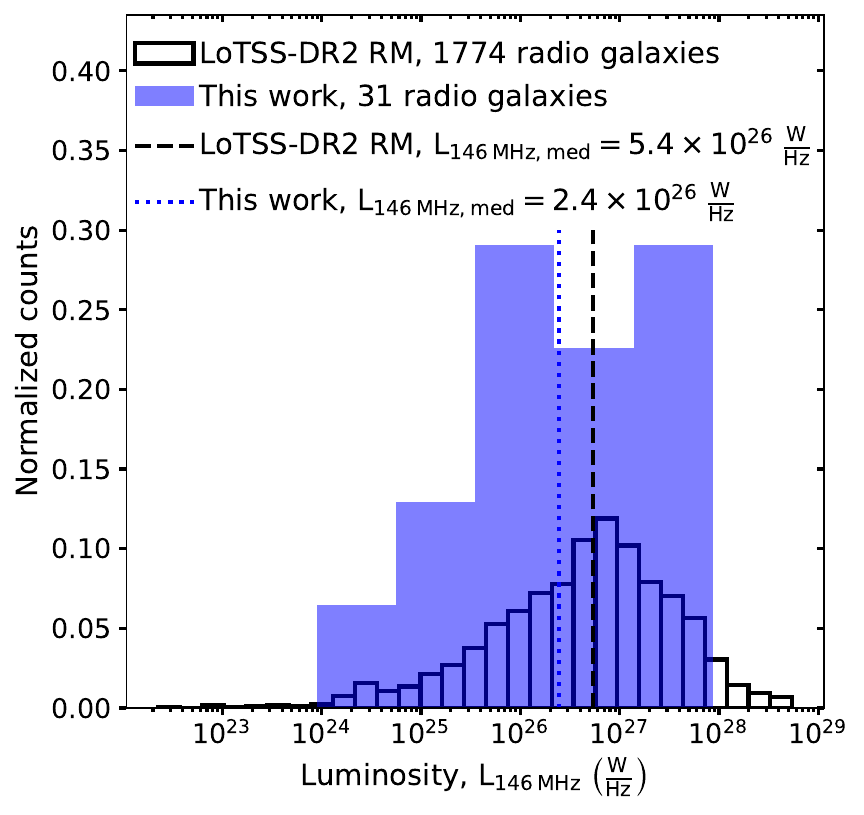} 
\caption{
Comparison of some properties of the ELAIS-N1 sample (this work) and the LoTSS-DR2 RM sample.
{\it Left panel:} Redshift distribution of the host galaxies of the radio galaxies in ELAIS-N1 in which we detected polarization (blue) 
and of the radio galaxies in the LoTSS-DR2 RM sample (black). 
{\it Right panel:} Distribution of rest-frame luminosities for our catalog (blue) and for the LoTSS-DR2 RM catalog computed at 146~MHz (black). The median values are indicated by vertical lines and given in Table~\ref{table:lumin}.
}
\label{fig:histozhistolum} 
\end{figure*}

\cite{HerreraRuiz2021} detected polarization in ten radio galaxies, including three compact ones, three FRIs, and  three FRII. The authors did not assign any morphological class to their source~2 (our source~7, which we classify as an FRII).
In the LoTSS Preliminary Data Release (at the low resolution of $4.3'$), \cite{VanEck2018} found only one FRI radio galaxy out of 92 polarized sources, which led \cite{HerreraRuiz2021} to suggest that deeper observations could enhance the detection rate of polarized emission from the extended regions of FRI radio galaxies.  
However, after stacking 19 epochs and considering Poissonian errors, our fraction of polarized FRI radio galaxies ($\sim$13\%) is lower than that of \cite{HerreraRuiz2021} (who reported 30\%). In all the FRIs that we detected, polarization was found in the core and not in the lobes. 

\subsection{Host galaxies and redshifts}

Table~\ref{table:catalog} lists 33 entries that correspond to polarized components associated with 31 radio galaxies (source~4 has two polarized components, 04$_{\rm A}$ and 04$_{\rm B}$, and components $13_{\rm A}$ and 29 ($13_{\rm B}$) belong to the same radio galaxy). As described in more detail below, we were able to obtain
redshift estimates from the literature for all radio galaxies in the sample ({\textcolor{black}{26}} spectroscopic and {\textcolor{black} 5} photometric). 

For about half of our sample, host galaxies and redshifts were reported by \cite{Simonte2024A&A...686A..21S} in their investigation of extended radio galaxies in the three LOFAR Deep Fields (which included the 25 deg$^2$ area of ELAIS-N1). 
A cross-match of our catalog with theirs provided a host identification and redshift for 16 radio galaxies (associated with our polarized components~4, 5, 6, 8, 11, 12, 13, 14, 16, 17, 18, 19, 22, 27, 31, and 32). 
For the remaining sources, we associated the host galaxy and the redshift from the catalog from the 16th data release from the Sloan Digital Sky Survey (SDSS) by \cite{Ahumada2020ApJS..249....3A}, using the images from the Dark Energy Spectroscopic Instrument (DESI) Legacy Imaging Surveys\footnote{\url{https://www.legacysurvey.org/}} \citep{DESI2024AJ....168...58D} and the NASA/IPAC Extragalactic Database (NED\footnote{\url{https://ned.ipac.caltech.edu/}}).

In Fig.~\ref{fig:histozhistolum} (left panel) we show the redshift distributions of the host galaxies of radio galaxies with at least one polarized component in ELAIS-N1 and of those in the LoTSS-DR2 RM catalog.
The median redshift is higher for the ELAIS-N1 catalog  than for the LoTSS-DR2 RM sample, as indicated in the legend.

\subsection{Rest-frame spectral luminosities} \label{sect:luminosities} 

The central frequency of the LOFAR observations of the ELAIS-N1 Deep Field is 146~MHz, and that of LoTSS-DR2 is 144~MHz.  
For the radio galaxies in our ELAIS-N1 catalog, we calculated the rest-frame spectral luminosities at 146~MHz following the equation \citep{Condon2018}
\begin{equation}
    L_{146~{\rm MHz}}= S_{146~{\rm MHz}} \, 
    \frac{4 \pi D_{\rm L}^2(z)} {(1+z)^{1 + \alpha}}\, ,
\end{equation}
where $S_{146~\rm{MHz}}$ is the flux density of the source, 
$D_{\rm L}$ is the luminosity distance, 
and $\alpha = -0.7$ is the spectral index ($S \propto \nu^{\alpha}$).  
In LoTSS-DR2, rest-frame luminosities at 144~MHz were derived using the same values of the cosmological parameters and assuming the same spectral index value. To compare the two samples, we scaled the published LoTSS-DR2 144~MHz rest-frame luminosities down by a factor of $(146/144)^{-0.7}\sim 0.99$ to obtain the rest-frame luminosities at 146~MHz. 

In ELAIS-N1, we estimated the flux densities of the sources from the Stokes $I$ image of \cite{Sabater2021} using Martin Hardcastle's code \texttt{radioflux}\footnote{\url{https://github.com/mhardcastle/radioflux}}, which sums the  intensities within a selected area of the image. For the sources in common with the catalog of \cite{Simonte2024A&A...686A..21S} (the radio galaxies associated with our polarized components~11, 12, 13, 15, 23, and 26), we used the values provided by those authors who measured the flux densities more carefully for these extended sources. 

\begin{table}[t]
\small
\caption{Median quantities of the ELAIS-N1 sample and of the LoTSS-DR2 RM sample of radio galaxies with a known redshift and at least one polarized component: 
redshift, flux density at 146~MHz, rest-frame 146 MHz spectral luminosity, or number of sources in each sample. 
}
\centering                         
\label{table:lumin}      
\begin{tabular}{ l c  c  c c}        
\hline\hline  
       & $z$    &  $S_{146~{\rm MHz}}$   & $L_{146~{\rm MHz}}$ &$N$ \\ 
       &        & (mJy)                 & (W~Hz$^{-1}$) \\ 
\hline
This work       & 0.71    & 190        & $2.4\times 10^{26}$ & 31\\ 
LoTSS-DR2 RM    & 0.61    & 403         & $5.4\times 10^{26}$ & 1774 \\ 
\hline
\end{tabular}
\end{table}

In Fig.~\ref{fig:histozhistolum} we show the distributions of 146~MHz rest-frame luminosities for the two samples. 
The median quantities are given in Table~\ref{table:lumin}. 
In ELAIS-N1 we detect polarization from radio sources that are fainter on average than those in the LoTSS-DR2 RM catalog (we scaled the 144~MHz flux densities from the LoTSS-DR2 RM catalog by the factor mentioned above before we calculated the median at 146~MHz). The population of polarized sources in ELAIS-N1 is at a higher median redshift than the polarized sources in LoTSS-DR2 RM catalog, and the median rest-frame luminosity is lower in ELAIS-N1. 
This was to be expected from deeper observations: We are able to detect some fainter sources that are more distant and/or intrinsically less luminous.  
As expected from the lower median rest-frame luminosity in ELAIS-N1, the fraction of sources above the traditional FRI/FRII luminosity separation of $\sim 10^{26}$~W~Hz$^{-1}$ is lower in ELAIS-N1 ($\sim 60$\% of the sources) than in the LoTSS-DR2 RM catalog (about 66\%).

\subsection{Linear sizes} \label{sect:linearsize}

\begin{figure} [h] 
\centering
\includegraphics[width=0.84\linewidth]{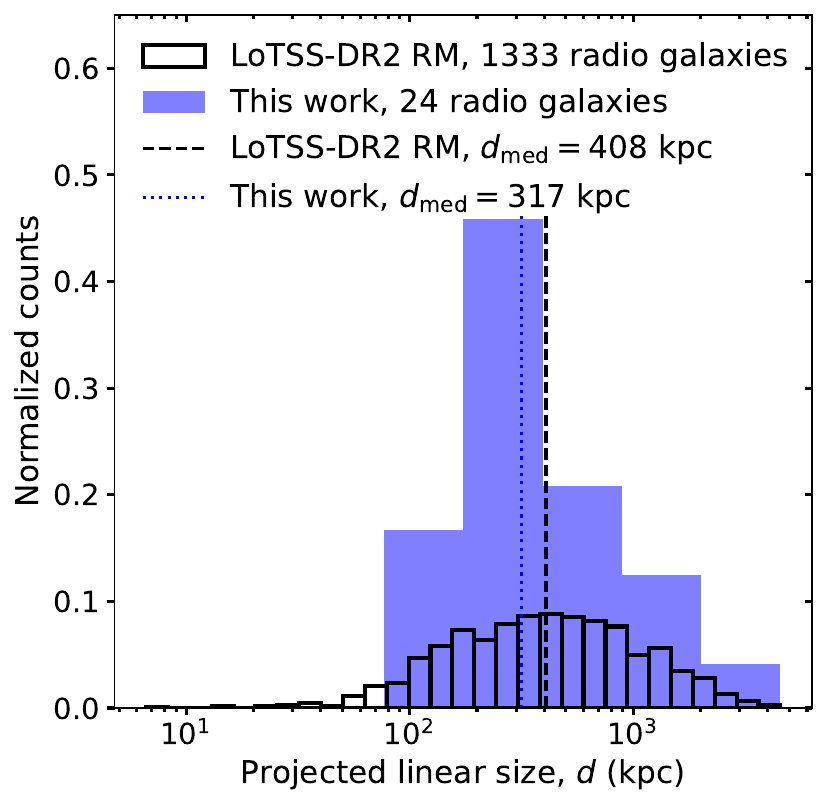} 
\caption{Distribution of the projected linear sizes of the subset of resolved radio galaxies in both catalogs. The median values are indicated.
}
\label{fig:histoLinSizes} 
\end{figure}

In Fig.~\ref{fig:histoLinSizes} we show the distribution of projected linear sizes of resolved radio galaxies associated with at least one polarized component in ELAIS-N1 and in the LoTSS-DR2 RM catalog. 
Radio galaxies in our samples are classified as resolved when their angular size is greater than 15$''$ (and as unresolved otherwise). This criterion is similar to the one used by \citep{Shimwell2022} and adopted for the LoTSS-DR2 RM catalog. In Table~\ref{table:catalog} we list the sizes (both the angular sizes and the projected linear sizes on the sky) of the radio galaxies in our sample. The projected linear sizes are always lower limits to the true sizes in three-dimensional space of the radio galaxies.

For 23 of the radio galaxies in which we detected polarization\footnote{those containing the polarized sources~02, 04, 05, 06, 08, 11, 12, 13, 14, 15, 16, 17, 18, 19, 20, 21, 22, 23, 26, 27, 30, 31, and 32},
we took the values of the angular sizes provided by \cite{Simonte2024A&A...686A..21S}. 
For 6 radio galaxies,\footnote{polarized sources~03, 09, 10, 24, 25, and 28} we took the values from \cite{Sabater2021}. 
For 2 radio galaxies (sources~1 and 7), we determined the angular sizes by inspecting the total-intensity image of \cite{Sabater2021}; this was necessary as these sources are FRII radio galaxies but were not resolved into two components in the catalog of \cite{Sabater2021}.

In the LoTSS-DR2 RM catalog, the majority of the radio galaxies in which polarization was found are large radio galaxies with a median projected size of $\sim$408~kpc for the resolved sources and $\sim$300~kpc (including upper limits)
for the unresolved sources.
The median projected linear size of our resolved radio galaxies is $\sim$317~kpc, and it is $\sim$235~kpc 
for the entire sample (including upper limits for the unresolved radio galaxies), suggesting that the radio galaxies in ELAIS-N1 in which we detect polarization with LOFAR are also mainly large radio galaxies. 
We find nine sources with a projected linear size greater than 500~kpc (Table~\ref{table:summaryResults}); three of these are giant radio galaxies (sources~11, 12, and 23) with sizes greater than 1~Mpc. 

\begin{figure*}
\centering
\includegraphics[width=0.45\linewidth]{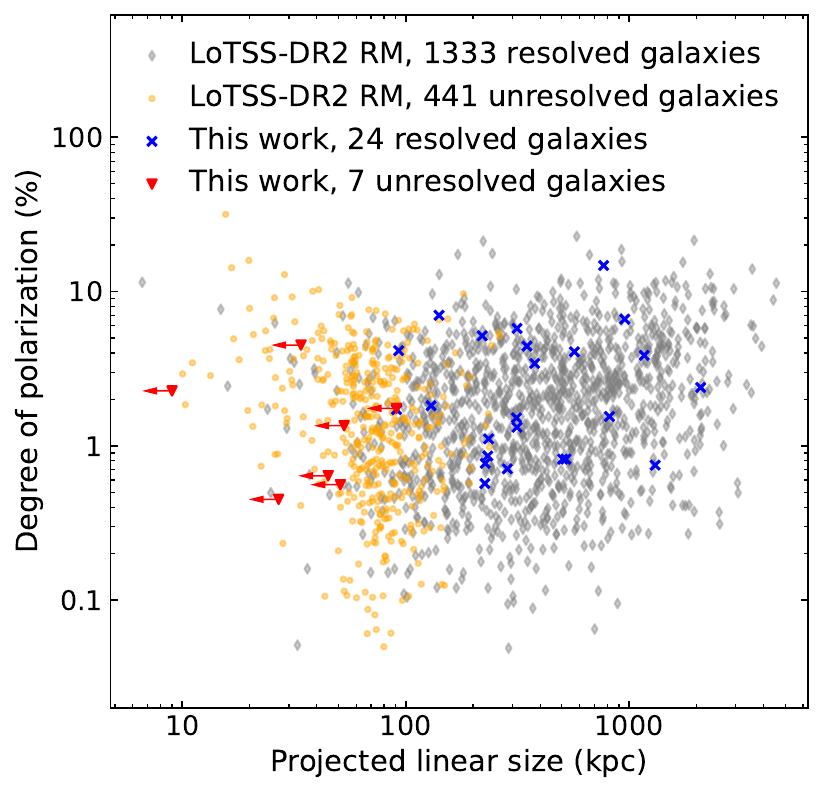}
\hspace{1cm}
\includegraphics[width=0.45\linewidth]{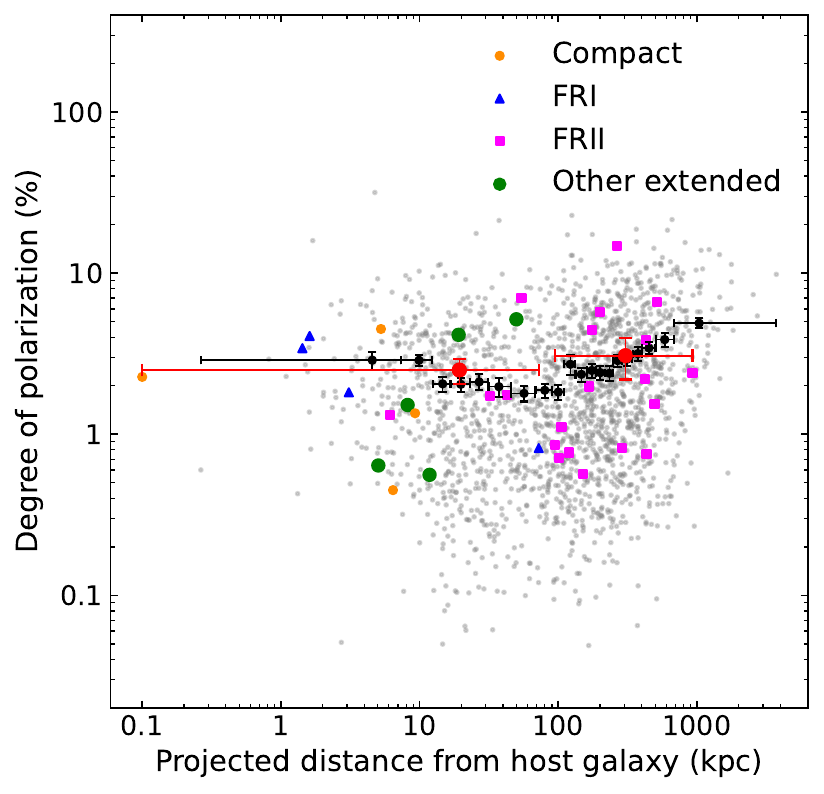}
\caption{
Comparison of the variations of the degrees of polarization with linear size and with projected distance from the host in the ELAIS-N1 sample (this work) and in the LoTSS-DR2 RM sample.
{\it Left panel:} Degree of polarization of the polarized component as a function of the projected linear source size 
for the radio galaxies in the LoTSS-DR2 RM sample and for those of our ELAIS-N1 sample, with upper limits on the linear source sizes for unresolved radio galaxies (shown in orange for the LoTSS-DR2 RM sample and in red with horizontal arrows for the ELAIS-N1 sample). 
{\it Right panel:} Degree of polarization of the polarized component as a function of the projected distance from the center of the host galaxy to the polarized component. 
The gray dots show the LoTSS-DR2 RM sample, and the other symbols represent different classes of radio galaxies in ELAIS-N1 in which polarization was detected.
The black dots show the mean values in bins of 100 values for LoTSS-DR2-RM, and the red dots are the values for ELAIS-N1, where two bins of approximately equal numbers of polarized components were used. The horizontal lines show the width of each bin, and the vertical lines show the standard errors on the mean values.
} 
\label{fig:pol_size} 
\end{figure*}

In Fig.~\ref{fig:pol_size} (left panel) we show the degree of polarization of the polarized component as a function of the projected linear size for the radio galaxies in which we detected polarization and for the LoTSS-DR2 RM sample. 
In both catalogs, the fractional polarization was computed at the pixel of the detection of polarization in the polarized intensity map and at the corresponding pixel in the total-intensity map.

In the LoTSS-DR2 RM catalog the degree of polarization increases with projected linear size. 
In ELAIS-N1 we considered the resolved radio galaxies and divided the sample into two bins of linear sizes, with equal numbers of components in each bin. 
We found a median degree of polarization of 
2.2\% for sizes in the 91--315~kpc range 
and of 4.1\% in the 315~kpc--2.1 Mpc size range.
While our findings are consistent with the increase in the degree of polarization observed in the LoTSS-DR2 RM catalog, they are limited by the small size of our sample (24 radio galaxies, 12 per bin).

In Fig.~\ref{fig:pol_size} (right panel) we show the degree of polarization of the polarized component as a function of projected distance from the center of the host galaxy to the polarized component for the 33 polarized components of the ELAIS-N1 LOFAR Deep Field 
and for the 1948 polarized components associated with host galaxies with a known redshift in the LoTSS-DR2 RM catalog. 
At projected distances from the host of $d > 100$~kpc, the degree of polarization of the polarized component increases with increasing distance from the host galaxy. 
At these projected distances, the polarized components are likely to be exclusively distant hotspots of FRII radio galaxies that are intrinsically strongly polarized; the fractional polarization continues to increase with distance from the host because the effect of the environment is weaker: Giant radio galaxies mostly reach large sizes because they expand in low-density environments; when the environment is sparser, the radio galaxy can become larger, and the distant hotspots/lobes are not as strongly depolarized. 

\section{LOFAR RM and RRM grid of ELAIS-N1} \label{sect:RM_RRMgrid}

The extragalactic RM, called residual rotation measure (RRM), is obtained by subtracting the foreground Galactic RM (GRM) from the observed RM, 
\begin{equation}
{\rm RRM} = {\rm RM} - {\rm GRM} \,. 
\end{equation}

To obtain the GRM values, we used the reconstruction of the Galactic Faraday depth sky of \cite{Hutschenreuter2022} and calculated the GRM as the average value within a circle with a diameter of one degree centered at the position of each RM source, as this is the typical separation between sources in the study by \cite{Hutschenreuter2022} (a similar approach was used by \citealt{Carretti2022} and \citealt{lotssdr2rm}). The uncertainties associated with the RRM values were computed as $\sigma_{\rm{RRM}}= \sqrt{\sigma_{\rm{RM}}^2+\sigma_{\rm{GRM}}^2}$, where $\sigma_{\rm{GRM}}$ is the mean value of the GRM uncertainty map from \cite{Hutschenreuter2022} within the circle with a diameter of one degree. 
The RM, GRM, and RRM values are reported in Table~\ref{table:catalog}. 

\begin{figure}[h]
\centering
\includegraphics[width=0.7\linewidth]{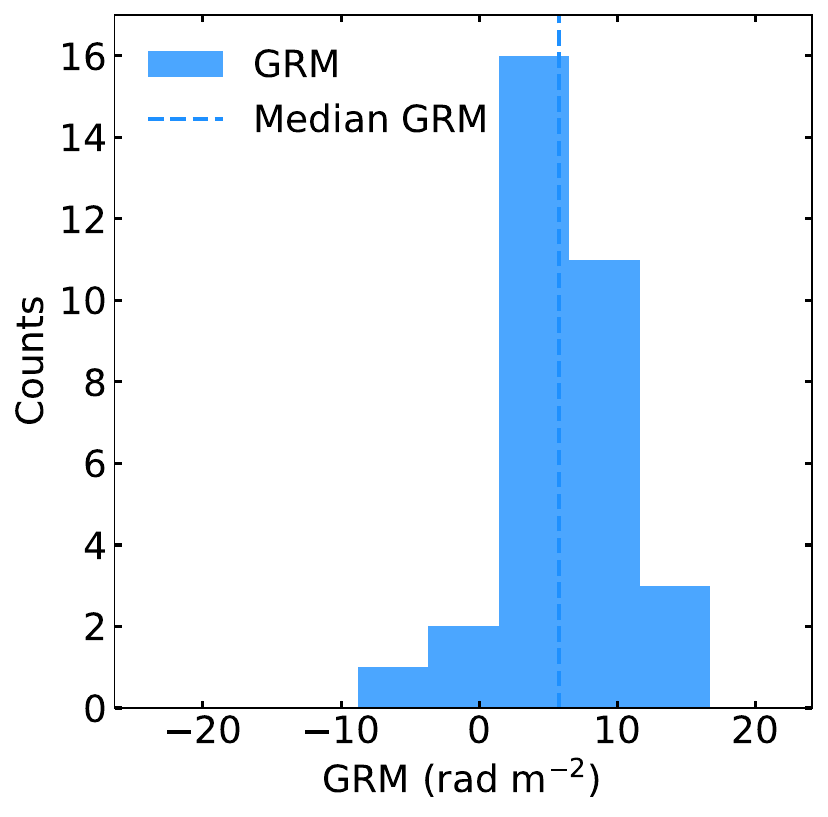}\\ 
\includegraphics[width=0.7\linewidth]{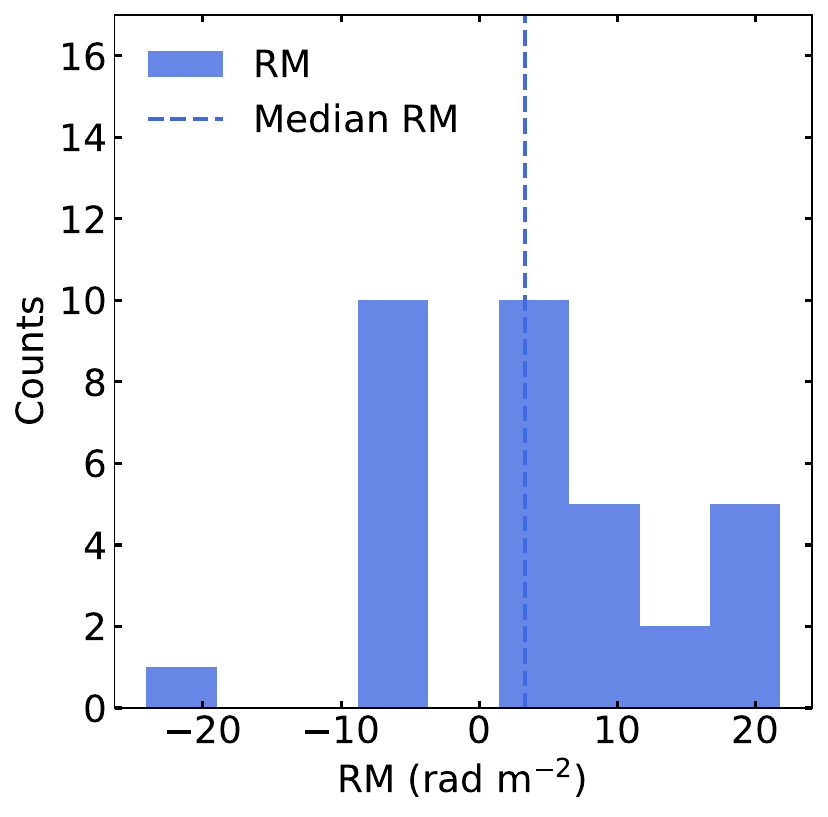}\\ 
\includegraphics[width=0.7\linewidth]{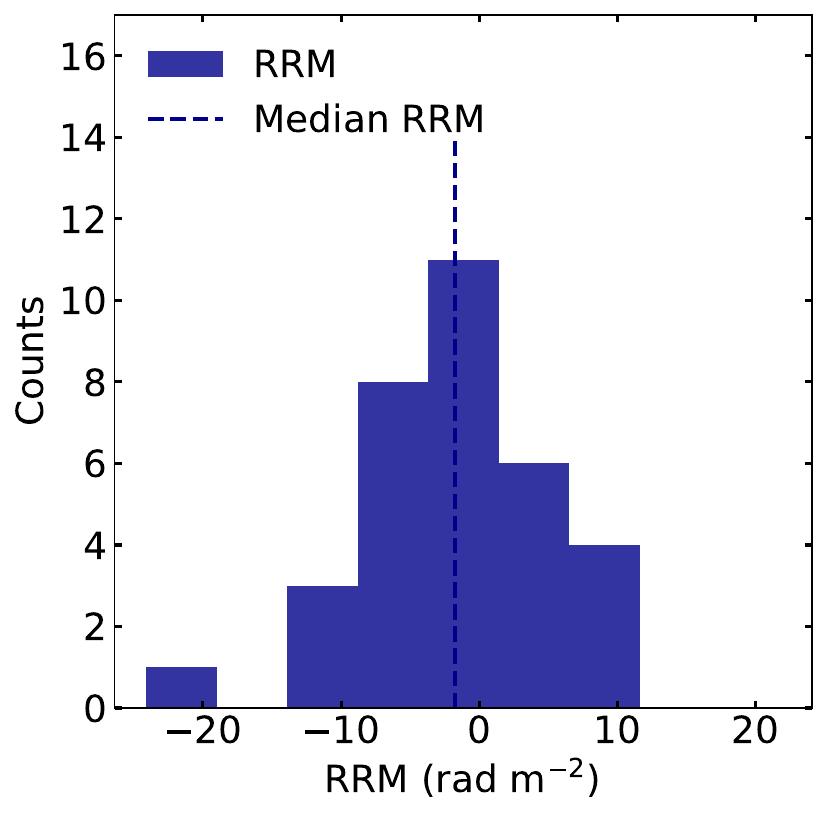}\\
\caption{Histograms of the GRM, RM, and RRM distributions for the 33 polarized components in the ELAIS-N1 LOFAR Deep Field.}
\label{fig:histoRMandRRM} 
\end{figure}

In Fig.~\ref{fig:histoRMandRRM} we show histograms of the RM, GRM, and RRM distributions. 
Because of instrumental polarization due to leakage from Stokes $I$ into Stokes $Q$ and $U$ (Sect.~3.4 of \citetalias{piras2024a}), there is a gap in the RM distribution around zero that is absent in the RRM distribution that is obtained after the Galactic RM values were subtracted. The GRM values estimated from the map of \cite{Hutschenreuter2022}  are mostly positive in the ELAIS-N1 field, with a mean value of about 6~rad~m$^{-2}$, and a scatter that is smaller than that in the RM distribution.  
The RRM distribution obtained after the GRM values were subtracted is narrower than the RM distribution   
The mean and the median RRM values are consistent with zero, as expected for a population of extragalactic sources whose polarized emission crosses the intergalactic magneto-ionic medium with random fluctuations of the electron-density-weighted parallel component of the magnetic field around zero. 
This was also reported in the LoTSS-DR2 RM studies of \cite{Carretti2022} and \cite{lotssdr2rm}. 

Some statistical values are given in Table~\ref{table:RRMstats}. 
In addition to the means and medians (and the dispersions around their values), we list in the two last columns 
two indicators of the strengths of the various contributions: 
the mean of the absolute values, $\langle {\rm X} \rangle$, 
and the root mean square (rms), $\langle \mathrm{X}^2 \rangle ^{1/2}$, 
where ${\rm X}$ is one of the RM-related variables (RM, GRM, or RRM). 
The rms is the most physically meaningful indicator because its square is related to the power ($\propto B_\parallel^2$) that is contained in the line-of-sight component of the magnetic field. 
The Galactic RM and the extragalatic RM (estimated by RRM) have comparable contributions, as estimated from their rms values ($\sim 7$~rad~m$^{-2}$). 

\begin{table}[ht] 
\small
\caption{Basic statistical properties of the RM and RRM distributions, all in rad~m$^{-2}$, of the 33 polarized components in the ELAIS-N1 LOFAR Deep Field.
} 
\centering                         
\label{table:RRMstats}      
\begin{tabular}{l r r r r c r}        
\hline\hline  
    &Mean  &std$^{a)}$ & Median & MAD$^{b)}$ & Mean(abs)$^{c)}$ & rms$^{d)}$\\ 
\hline
RM  & 4.21  & 9.56  & 3.29  & 7.30    &8.48     &10.44\\ 
GRM & 6.13  & 4.09  &5.74   & 3.32    &6.37     &7.37\\
RRM &-1.92  & 6.77  &-1.78  & 4.51    &5.45     &7.03\\ 
\hline
\end{tabular}
\tablefoot{
a) Standard deviation. The quoted values use the default of the {\tt numpy.std} Python command, which uses a normalization by $N$ rather than by $N-1$.
b) Median absolute deviation.
c) Mean of the absolute values.
d) Root mean square.
}
\end{table}

\begin{table}
\centering
\small
\caption{Spearman correlation coefficients and $p$ values for the sample of polarized components in the LOFAR ELAIS-N1 Deep Field.} 
\label{table:correl}    
\begin{tabular}{l c c}  
\hline\hline  
        &Spearman coefficient  &$p$-value\\ 
\hline
$\Pi$ vs $z$    &$-0.24$  &0.19\\
RRM vs $z$      &$-0.09$  &0.60\\
RRM vs $\Pi$    &$+0.02$  &0.89\\
\hline
\end{tabular}
\tablefoot{$\Pi$ is the degree of polarization.}
\end{table}

In their reconstruction of the GRM sky, \cite{Hutschenreuter2022} used a number of RM catalogs, the largest of which were the NVSS RM catalog of \cite{taylor2009} and the LoTSS-DR2 RM catalog of \cite{lotssdr2rm}. 
In the 25 deg$^2$ ELAIS-N1 LOFAR Deep Field, the NVSS catalog has 27 RM entries (2 of which are associated with sources that lie very close to the northern border of the field and were not included in the accounting presented in Table~1 of \citetalias{piras2024a}); the LoTSS-DR2 RM catalog has 9 entries (8 of these polarized sources are in our catalog; the ninth source is in the outskirts of our LOFAR pointing and was detected in LoTSS-DR2 RM because of the better signal-to-noise ratio that was gained by the mosaicking of two pointings at this location). 

\begin{figure}[h!] 
\includegraphics[width=0.7\linewidth]{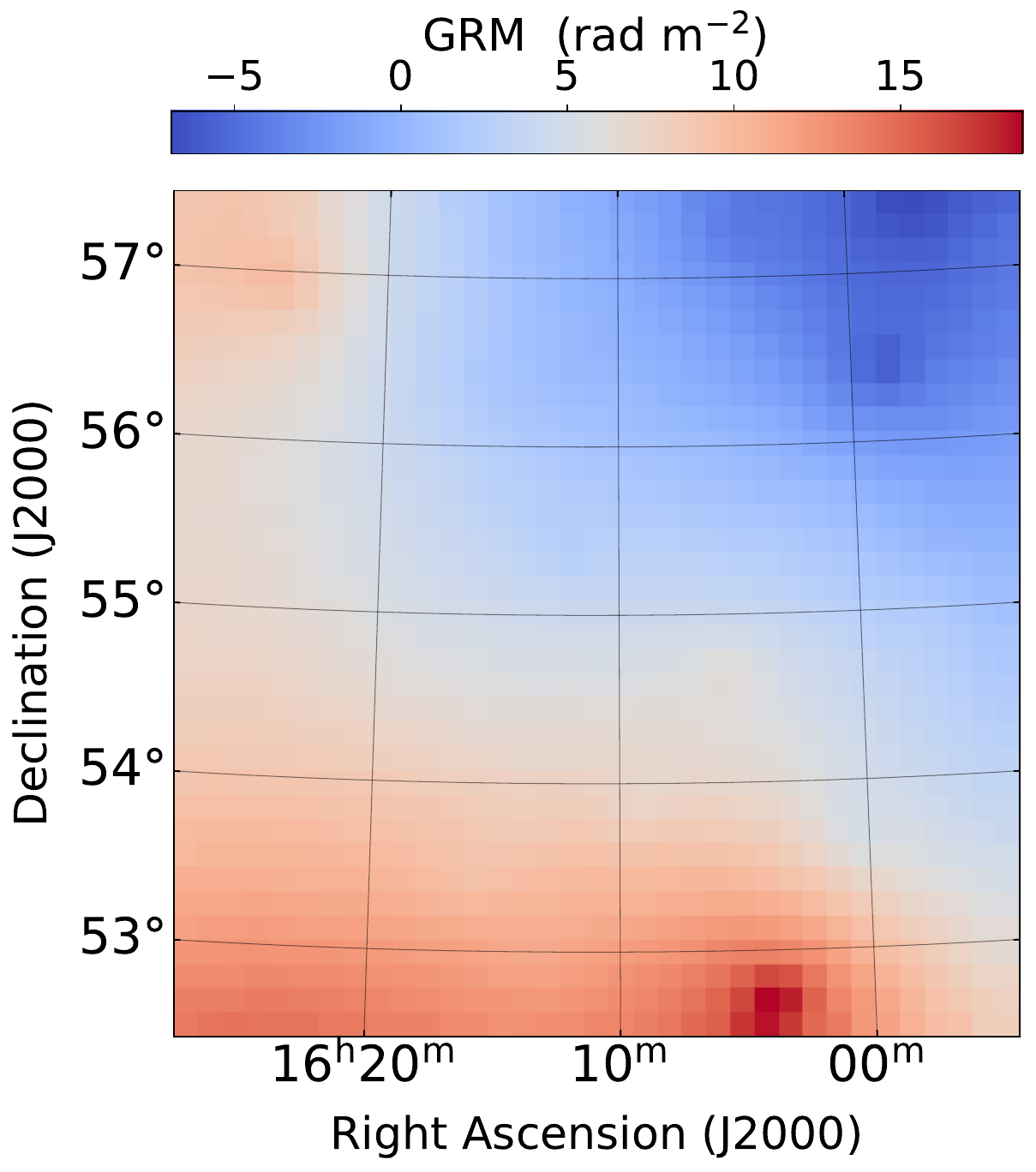} 
\includegraphics[width=0.7\linewidth]{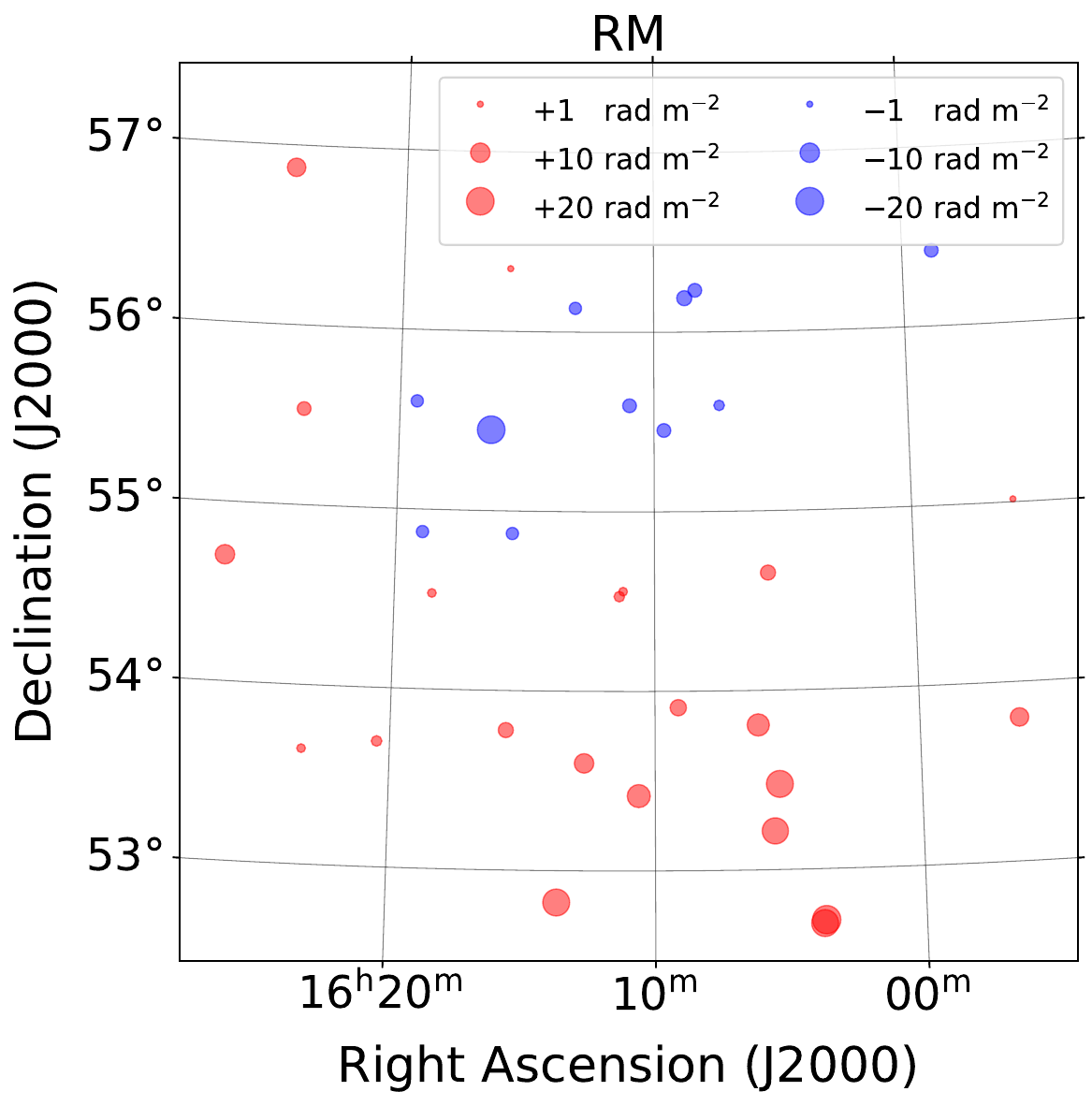}
\includegraphics[width=0.7\linewidth]{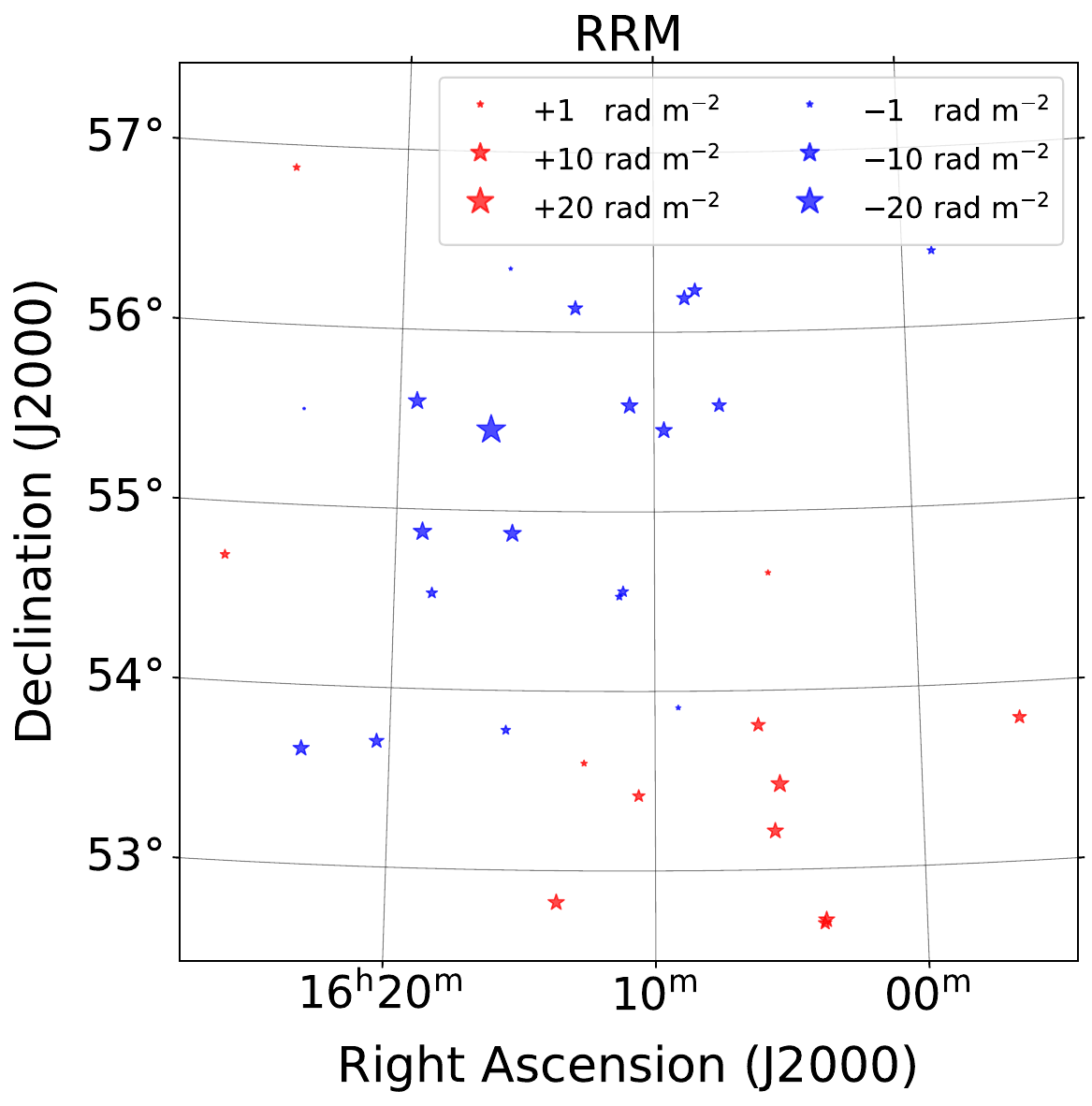}
\caption{Images of the GRM, RM, and RRM values.
{\it Top panel:} Galactic RM map from \cite{Hutschenreuter2022}. 
{\it Other panels:} Distribution of polarized sources in the ELAIS-N1 LOFAR Deep Field. 
The size of the markers is proportional to the magnitude of the RM values
(middle panel) and RRM values (bottom panel).}
\label{fig:RMgrid} 
\end{figure}

In Fig.~\ref{fig:RMgrid} (first panel) we show the GRM map of the 25 deg$^2$ field of ELAIS-N1 extracted from the all-sky Galactic RM map of \cite{Hutschenreuter2022}. 
The clear gradient across the field is also visible in the polarized foreground 
imaged with LOFAR at a resolution of $4.3'$ by \cite{Snidaric2023}. 
The other panels show the RM grid and the RRM grid obtained after subtraction of the GRM values. 
While the polarized foregrounds may originate in nearby Galactic structures and probably do not probe the full line of sight through the Milky Way, the similarity between the two maps is an indicator that the Galactic RM map of \cite{Hutschenreuter2022} is a reasonable estimate of the GRM. However, the fact that the gradient seen in the GRM map remains in the RRM grid might be a sign of a residual GRM contamination.  

\subsection{Redshift dependence of the RRM and fractional polarization}

We examined the behaviors of the degree of polarization and of the absolute values of RRM, |RRM|, as a function of redshift, 
and of |RRM| as a function of degree of polarization. 
The exploration of these quantities can give us clues about the mechanisms that cause depolarization at different frequencies. As a statistical indicator of a possible correlation, we used the Spearman rank correlation coefficient (a value close to zero indicates no correlation; a value close to $+1$ indicates a correlation, and $-1$ indicates an anticorrelation). 
We also used the probability ($p$) value, which gives the probability that the null hypothesis (the absence of a correlation between the two considered variables) is correct and that an observed correlation is due to chance. 
The lower the $p$ value, the higher the probability that the correlation is real.
The numbers are given in Table~\ref{table:correl}.

\subsubsection{Degree of polarization versus redshift}

Of the three combinations of variables that we considered, the combination with the strongest Spearman correlation coefficient and lowest $p$ value is between the degree of polarization and redshift. These numbers indicate a weak anticorrelation.
In the left panel of Fig.~\ref{fig:fracpolVSz}, we show the fit obtained by \cite{Carretti2022} for the LoTSS-DR2 RM catalog in addition to the values for the 31 polarized components in ELAIS-N1. The anticorrelation in their catalog is stronger. 
\cite{Berger2021} also reported a decrease in the degree of polarization with increasing redshift in their 1.4~GHz observations of the Lockman Hole.

\subsubsection{RRM versus redshift}

Studying the NVSS RM catalog, \citep{Hammond2012arXiv1209.1438H} found no significant evolution of the RRM with redshift. 
At 144~MHz, \cite{Carretti2022} reported no variation in the RRM with redshift either. 
The central panel of Figure~\ref{fig:fracpolVSz} shows the absolute values of the RRM as a function of redshift. 
We do not find any sign of a correlation.

\subsubsection{RRM versus degree of polarization}

Finally, we examined the possible variations in the RRM with the degree of polarization of the polarized components. 
In the LoTSS-DR2 RM sample, the RRM appears to be independent of the degree of polarization, which contradicts the findings at 1.4~GHz of \cite{Hammond2012arXiv1209.1438H}, who reported a strong anticorrelation between the RRM and the degree of polarization in the NVSS RM data. \cite{Carretti2022} attributed this discrepancy to the different astrophysical origins of the RRM: 
Depolarization at 1.4~GHz would primarily stem from the environment of the local source, 
while at 144~MHz, it would predominantly be the result of radiation traveling through the intergalactic medium on large scales.

\begin{figure*}[h]
\centering
\includegraphics[width=0.33\linewidth]{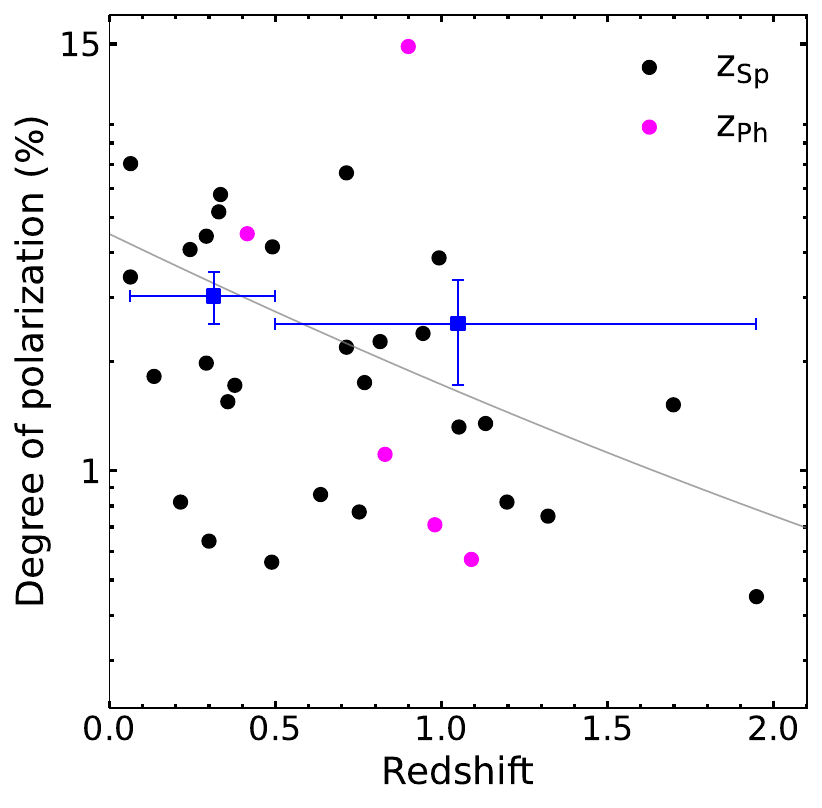} 
\includegraphics[width=0.33\linewidth]{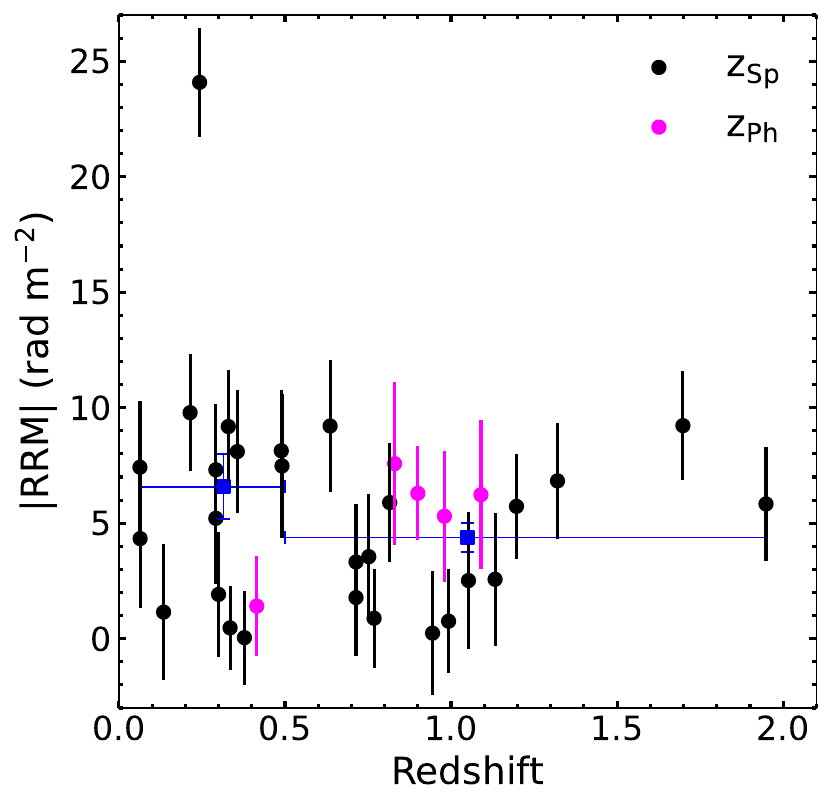} 
\includegraphics[width=0.33\linewidth]{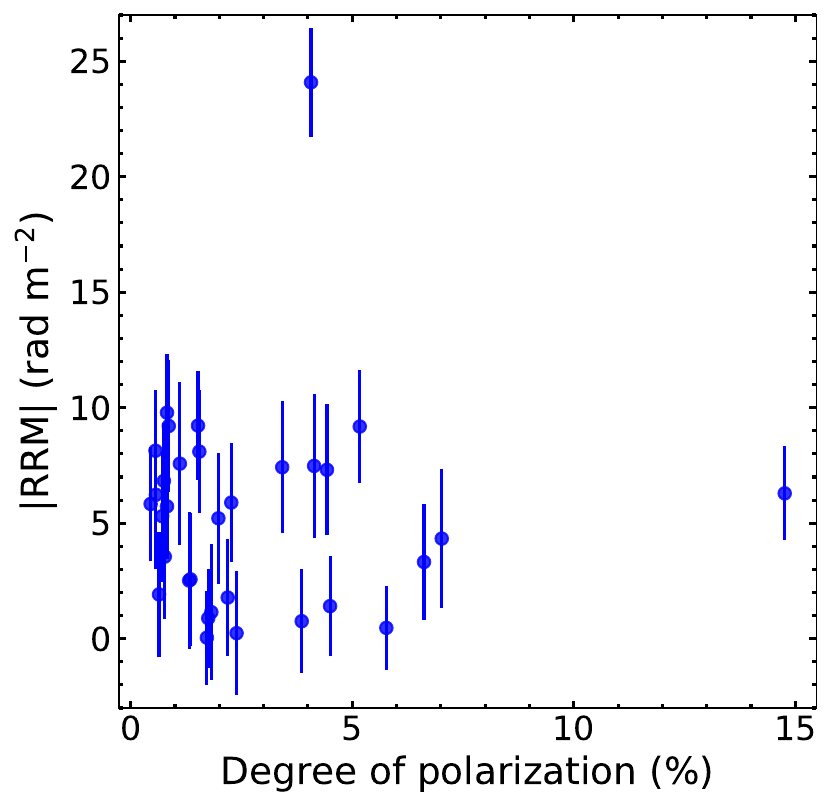} 
\caption{
Variations of degrees of polarization, $RRM$ and redshift values.  
{\it Left and middle panels:} Degree of polarization and |RRM| as a function of redshift. 
The sample was split to have a roughly equal number of sources in each redshift bin.  
Data points from sources with spectroscopic or photometric redshifts are indicated in different colors, as indicated in the legend.  
The horizontal bars indicate the sizes of the redshift bins, and the vertical bars show the standard errors of the values in the vertical axis. 
For each plot, the blue squares represent the means of the values in each bin. 
In the left panel, the gray line is the fit from \cite{Carretti2022} to corresponding data from the LoTSS-DR2 RM grid catalog. 
{\it Right panel:} |RRM| vs. degree of polarization.
}
\label{fig:fracpolVSz} 
\end{figure*}

The right panel of Fig.~\ref{fig:fracpolVSz}
shows the absolute RRM values as a function of the degree of polarization in the ELAIS-N1 Deep Field; these quantities seem to be independent, in agreement with the findings of \cite{lotssdr2rm} and \cite{Carretti2022}.

\subsection{Polarized sources in relation to the large-scale structure}

The investigation of the extragalactic environment offers information about the magneto-ionic media that interact with the polarized radiation.
There is observational evidence for the presence of magnetic fields within galaxy clusters (e.g., \citealt{Govoni2004, Clarke2004}) and intergalactic filaments (e.g., \citealt{O'Sullivan2019A&A...622A..16O}). 
Polarized radio sources in the background of clusters or in the clusters themselves may provide information on intracluster magnetic fields in the cluster that causes the Faraday effect. 
At 1.4~GHz, \cite{Bonafede2011} and \cite{Osinga2022A&A...665A..71O} found that the fractional polarization of background sources decreases toward the cluster center. 
\cite{Stuardi2020} studied the intergalactic magnetic fields along the lines of sight of giant radio galaxies detected in the LoTSS-DR2 and showed that the detection of polarized sources is disfavored  by the presence of foreground galaxy clusters. They found that the higher RM variance due to turbulence in a foreground intracluster medium influences the fractional polarization at gigahertz frequencies and depolarizes the radio emission at 144~MHz below the detection limit of LoTSS. Furthermore, they suggested that the background polarized sources can be detected probably only under particular conditions, for example, when the foreground cluster is poor and/or the polarized emission originates in a very compact region.  

\begin{table*}[ht]
\caption{Nearby superclusters in the field and their cluster members.}
\centering
\label{table:superclusters}
\begin{tabular}{ c c c c c c c c} 
\hline\hline
Source &  RRM & $z_{\rm source}$ & Supercluster & $z_{\rm supercluster}$ &Cluster & RA, Dec, $z_{\rm cluster}$\\
 &     (rad~m$^{-2}$)          &          &       &     & &  (deg, deg) \\
 (1)             &(2)            &(3)        &(4) &(5)   &(6) &(7) \\
\hline
08 & $+5.73 \pm 2.26$ & \multicolumn{1}{c|} {1.1215} & MSCC~476$^{(a)}$, SCL~162   &0.065    &A2168                     &243.27, +54.16, 0.064\\
12 & $-0.76 \pm 2.23$ & \multicolumn{1}{c|} {0.9928}      &       &         &A2149A                   &240.35, +53.87, 0.0657\\ 
 \hline
&        &  & MSCC~473$^{(b)}$    &   0.141     &A2149E     &240.35, +53.87, 0.139\\
 &               &                  &       &         &A2149F     &240.35, +53.87, 0.142\\
\hline
\end{tabular}
\tablefoot{
(4) MSCC is the Main SuperCluster Catalog of \cite{ChowMartinez2014}. 
SCL: Supercluster in the \citealt{Einasto2001AJ....122.2222E} catalog. 
(6) A2149 consists of several components along the same line of sight, so that the coordinates of the components are the same as those of A2149. 
(7) The redshifts of the clusters are from \cite{ChowMartinez2014}. 
$^{(a)}$ Supercluster with J2000 coordinates 
RA = 241.84$^\circ$, 
Dec = 54.01$^\circ$. This position corresponds to the geometrical center between the two members of the supercluster, A2168 and A2149A. 
The three-dimensional separation between the two clusters in the supercluster is 10.8~$h^{-1}_{70}$Mpc \citep{ChowMartinez2014}.
$^{(b)}$: Supercluster with J2000 coordinates 
RA = 240.35$^\circ$, 
Dec = 53.87$^\circ$. 
The two clusters in this supercluster are in the same position on the sky, and their line-of-sight separation is  12~$h^{-1}_{70}$Mpc \citep{ChowMartinez2014}.
}
\end{table*}

We compared the positions of our polarized sources with those of the 1.58 million clusters of galaxies cataloged by \cite{Wen2024ApJS..272...39W}, who estimated photometric redshifts and collected available spectroscopic redshifts of galaxies up to $z\sim 1.5$.  
We also compared the positions of our polarized sources with the sample of galaxy clusters from \cite{Zou2022RAA....22f5001Z}, who estimated photometric redshifts for galaxies up to $z\sim1$.
Within 25 deg$^2$ of ELAIS-N1 lie 1186 galaxy clusters from the catalog of \cite{Wen2024ApJS..272...39W}, and 548 from the catalog of 
\cite{Zou2022RAA....22f5001Z}. The two catalogs have 125 galaxy clusters in common.

\begin{figure}[h!] 
\centering
\includegraphics[width=\linewidth]{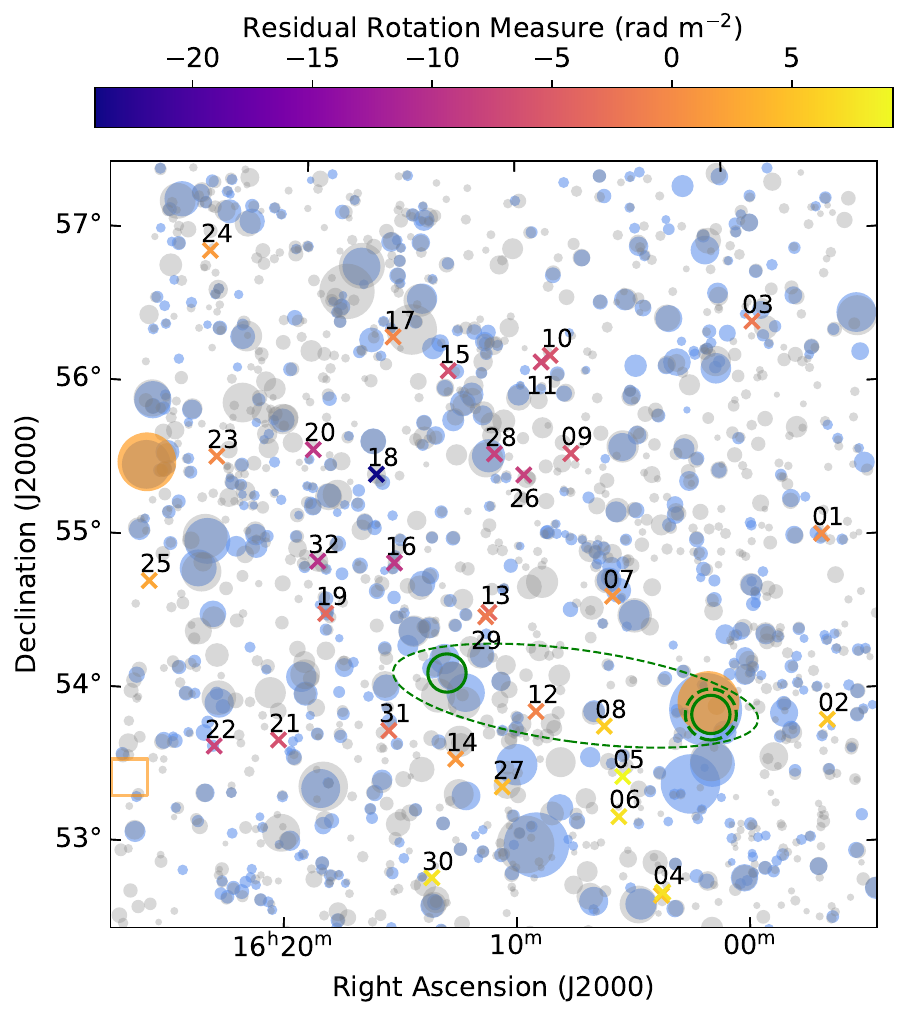} 
\caption{Galaxy clusters (shaded circles), superclusters, and polarized sources (crosses, with their RRM values as indicated in the color bar) in the ELAIS-N1 LOFAR Deep Field. 
The radii of the circles indicate $r_{200}$ of the clusters, 
except for the clusters that are located in superclusters, whose sizes are not known; we used an arbitrary diameter of $30'$ for the plot (solid green circles). 
The shaded gray circles represent the clusters of \cite{Wen2024ApJS..272...39W} 
and the shaded blue circles show the clusters from \cite{Zou2022RAA....22f5001Z}. 
The dashed green ellipse and the circle 
represent the two superclusters from \cite{ChowMartinez2014} in the field 
(Table~\ref{table:superclusters}).
The shaded orange circles represent the clusters detected in X-rays from \cite{Boringer2000ApJS..129..435B} and/or the PSZ2 \cite{PlanckCollaboration2016XA&A...594A..27P} catalogs. The orange square represents the X-ray Sunyaev-Zel’dovich cluster candidate from \cite{Tarrio2019A&A...626A...7T}, for which no radius is available.
}
\label{fig:clusters}
\end{figure}

Figure~\ref{fig:clusters} shows the locations of the polarized sources (as crosses) 
and of the galaxy clusters (as shaded circles).
The radius of each circle corresponds to $r_{200}$, the radius within which the mean density of a cluster is 200 times the critical density of the Universe at the cluster's redshift, and which approximately represents the outer boundary of a galaxy cluster. 
The catalogs provide $r_{500}$ values (in Mpc) for each galaxy cluster, 
and we converted this into angular sizes using the cosmological parameters adopted by the authors\footnote{\label{sizes} \cite{Wen2024ApJS..272...39W}, \cite{Zou2022RAA....22f5001Z}, and \cite{ChowMartinez2014} adopted the same values for the cosmological parameters, 
$H_0 = 70$~km~s$^{-1}$~Mpc$^{-1}$, $\Omega_{\rm m} = 0.3$, and $\Omega_\Lambda = 0.7$
}. 
We estimated $r_{200}$ using the relation $r_{500}/r_{200} = 0.7$ \citep{Ettori2009A&A...496..343E}. 

We also show in Fig.~\ref{fig:clusters} the locations of the two nearby superclusters in the field that were identified by \cite{ChowMartinez2014}, who studied superclusters of the Abell/ACO galaxy clusters out to redshift 0.15.  
The properties of these structures are summarized in Table~\ref{table:superclusters}.
For each supercluster, \cite{ChowMartinez2014} gave the value $d_{\rm max}$, which is the maximum projected distance (in Mpc) between the two member clusters of a supercluster farthest from each other, as a rough indicator of the size of a supercluster. We converted these values into angular sizes using the values that they used for the cosmological parameters (given in footnote \ref{sizes}). 

\begin{table*}[h] 
\caption{Polarized sources in the ELAIS-N1 LOFAR Deep Field that might be located behind or are embedded within the galaxy clusters identified by \citealt{Zou2022RAA....22f5001Z} (Clustering by Fast Search and Find of Density Peaks; CFSFDP) and/or \citealt{Wen2024ApJS..272...39W} (WH).}
\centering                         
\label{table:clusters}      
\begin{tabular}{ c c c c c c c c c}        
\hline\hline  
    Source  & Source Redshift  & Cluster &  Cluster Redshift & $R_{200}^{a)}$ &$\theta_{\rm sep}^{b)}$ &$d_{\rm sep}^{c)}$\\ 
            &                   &           &  & (Mpc) &(arcsec)    & (Mpc)\\ 
\hline 
07     &  0.7683  & CFSFDP~464000084    & 0.2440 & 1.14  & 196  &  0.75 \\ 
10     &  1.9482  & WH~J160827.0+561505 & 0.5396 & 0.78  &85    &  0.54 \\ 
16$^*$ &  0.32936 & WH~J161531.1+545230 & 0.3294 & 0.72  &10    & 0.05 \\ 
18$^*$ &  0.24297 & WH~J161623.7+552702 & 0.2430 & 0.85  &0     & 0  \\ 
19     &  0.7523  & WH~J161832.0+543337 & 0.3918 & 0.88  &94    & 0.50 \\ 
       &          & WH~J161820.9+543040 & 0.7944 & 1.09  & 136  & 1.02 \\ 
22     &  0.4910  & CFSFDP~483700074    & 0.4672 & 0.84  &27    & 0.16 \\ 
26$^*$ &  0.3566  & WH~J160929.8+552542 & 0.2602 & 0.87  &108   & 0.43  \\ 
       &          & WH~J160919.5+552549 & 0.3572 & 0.74  &169   & 0.85 \\ 
28     &  0.48882 & CFSFDP~445000147    & 0.2234 & 1.40  & 146  & 0.52  \\ 
32$^*$ &  0.21431 & WH~J161858.7+545228 & 0.2139 & 0.86  &1     & 0.003 \\ 
\hline
\end{tabular}
\tablefoot{
Sources with an asterisk are in a cluster. Three of these (sources 16, 18, and 32) are associated with radio galaxies whose host galaxy is the brightest cluster galaxy.\\
a) $R_{200} = R_{500}/0.7$, calculated from the $R_{500}$ values given by the authors in their adopted cosmology (footnote \ref{sizes}). \\
b) Angular separation between the polarized source and the cluster center.\\
c) Projected linear separation between the polarized source and the cluster center. 
}
\end{table*}
 
We noted that none of the supercluster members are listed in the cluster catalogs of  \cite{Wen2024ApJS..272...39W} or \cite{Zou2022RAA....22f5001Z} at the redshift indicated by \cite{ChowMartinez2014}, and that cluster sizes were not provided by the latter. Therefore, we indicate their positions in Fig.~\ref{fig:clusters}, but their sizes are arbitrary ($30'$ diameter).
Two of our polarized components are positioned behind the supercluster MSCC~476 (sources~8 and 12).

The two superclusters from \cite{ChowMartinez2014} correspond to pairs of clusters, where one pair is a redshift component along the line of sight of Abell~2149 (the westernmost of the green circles in Fig.~\ref{fig:clusters}). Supercluster catalogs are very sensitive to the linking length that is used, and the physical reality of these superclusters is doubtful especially for very low numbers of member clusters in a supercluster (e.g., only two for both superclusters discussed here). 
\cite{Sankhyayan2023ApJ...958...62S} published a catalog of SDSS superclusters using the cluster catalog of \cite{Wen2015ApJ...807..178W}. 
This supercluster catalog does not contain the MSCC clusters 473 and 476 from \cite{ChowMartinez2014}. It does not contain any supercluster in the ELAIS-N1 region. This does not mean that there are none, but that it has been very difficult to identify these structures. 
The very recent cluster catalogs of \cite{Zou2022RAA....22f5001Z}, 
\cite{Wen2024ApJS..272...39W}, and 
\cite{Yantovski-Barth2024MNRAS.531.2285Y} 
could be used to produce a new large supercluster catalog, but this is beyond the scope of this paper. 

In Table~\ref{table:clusters} we list the nine polarized components that are located along the lines of sight of clusters and whose redshifts are higher than or comparable to those of these clusters. 
The field also contains galaxy clusters detected in the X-ray and/or through the Sunyaev-Zel’dovich effect \citep{PlanckCollaboration2016XA&A...594A..27P, 
Tarrio2019A&A...626A...7T, 
Boringer2000ApJS..129..435B}, but none of them is positioned along the line of sight of the detected polarized sources. 

To search for a possible excess in RRM due to interaction of radiation from polarized sources with the foreground large-scale structure, 
we computed the mean absolute value of RRM of the sample of polarized components located in a cluster and/or behind a cluster or a supercluster (11 components: the 9 components in Table~\ref{table:clusters}, and the 2 components in Table~\ref{table:superclusters})
in comparison to the value of polarized sources that are not along the line of sight of any structure (22 components). 
We used the standard error ($\sigma$) 
as a representation of the uncertainty on the values. 
We found 
$8.6 \pm 2.0$~rad~m$^{-2}$ and
$4.4 \pm 0.6$, respectively, which agree within $2\sigma$.
This suggests that the magnetic fields at that position of the clusters are relatively weak and have a low plasma density. Polarized radiation from background sources that pass through regions of the clusters with a high plasma density and strong magnetic fields, such as the cluster centers, may undergo depolarization due to the high RM dispersion (e.g., \citealt{Osinga2022A&A...665A..71O}) and be completely depolarized at LOFAR frequencies and sensitivities. Similar findings were reached at 1.4~GHz in the XXL-South field by \cite{Eyles2020}, who also detected polarized sources at the edge of the clusters, as defined by their $r_{200}$ values. 

Our sample contains four sources that are embedded within clusters (Table~\ref{table:clusters}). 
In one case (source~32), the distorted morphology of the radio galaxy is indeed suggestive of an interaction of the radio plume with a surrounding intracluster medium.  

The absolute RRM values of the sources behind the supercluster agree within $2\sigma$ with the mean absolute value of the sources that are not located behind any structure. Again, this may indicate very weak magnetic fields and low-density plasma at these positions of the supercluster. A larger sample is needed to measure an effect.

\section{Conclusions}\label{sec:conclusion}
We characterized the polarization properties of 31~sub-millijansky extragalactic sources observed in the ELAIS-N1 LOFAR Deep Field that were originally presented in \citetalias{piras2024a}. The work is based on the most sensitive polarization data acquired at 146~MHz to date.
Our analysis revealed that the morphological featuers of nearly half of the sources are consistent with FRII radio galaxies, and the remaining half is divided among FRI-type radio galaxies, compact sources, and diffuse sources. Notably, similarly to the LoTSS-DR2 RM catalog of \cite{OSullivan2020}, the submillijansky-polarized sources tend to be large radio galaxies. 
Our deeper observations enabled the detection of fainter sources that are more distant and/or intrinsically less luminous sources, as expected. 

We produced a residual rotation measure map of  the ELAIS-N1 region. The median RRM value is close to zero, as in the LoTSS-DR2 RM. This is 
an indication that the Galactic RM inferred by \cite{Hutschenreuter2022} was a reasonable estimate, and 
it suggests that the polarization angle experiences a random walk through the intervening intergalactic magneto-ionic medium. However, some of the overall gradient seen the GRM map remains in the RRM map, which may be a sign of residual GRM contamination. Future deep polarization surveys such as those planned with the Square Kilometre Array will enable the construction of denser RM grids and improved GRM/RRM separations. 

Furthermore, we found that the RRM values at 146~MHz remain independent of the degrees of polarization for submillijansky sources, which further supports the notion that the RRM at these frequencies is dominated by contributions from the intergalactic medium and not by the local medium. By plotting degrees of polarization versus projected distance of the polarized components to the centers of the host galaxies for the LoTSS-DR RM catalog, we found an increase beyond 100~kpc, which indicates that the local medium contributes to the depolarization. 

When we compared our source population to the known locations of clusters and superclusters from the existing literature, we observed that the depolarization and RM properties of sources toward clusters and superclusters are similar to those observed in the entire source population. Consequently, it appears that these structures contribute very little to these properties. 

Our work highlights the need for deeper surveys to reveal previously undetected populations of polarized extragalactic sources and to fully characterize their properties to understand the cosmic magnetic field structure.

\section{Data availability}
The full catalog of properties of detected polarized sources and references are available in electronic form at the CDS via anonymous ftp to cdsarc.u-strasbg.fr (130.79.128.5) or via \url{http://cdsweb.u-strasbg.fr/cgi-bin/qcat?J/A+A/}.
Appendix~\ref{app:sources_images} is available at the link \url{https://zenodo.org/records/14012234}. 

\begin{acknowledgements} 
We are very grateful to the referee, Heinz Andernach, for a constructive and thorough report that led to a significant improvement of the paper. 
LOFAR, the Low Frequency Array designed and constructed by ASTRON, has facilities in several countries, that are owned by various parties (each with their own funding sources), and that are collectively operated by the International LOFAR Telescope (ILT) foundation under a joint scientific policy. 
This work was done within the LOFAR Surveys and the LOFAR Magnetism Key Science Projects. 
This research has made use of the NASA/IPAC Extragalactic Database (NED) which is operated by the Jet Propulsion Laboratory, California Institute of Technology, under contract with the National Aeronautics and Space Administration.
This research has made use of the VizieR catalog access tool, CDS, Strasbourg, France. The original description of the VizieR service was published in A\&AS 143, 23.
We have also made use of the table analysis software {\tt topcat} \citep{topcat}.
This research made use of {\tt Astropy}, a community-developed core
Python package for astronomy \citep{astropy}, of {\tt Matplotlib} \citep{Hunter2007}, and of
{\tt APLpy} \citep{RobitailleBressert2012}, an open-source astronomical
plotting package for Python. 

This research has made use of the CIRADA cutout service at URL cutouts.cirada.ca, operated by the Canadian Initiative for Radio Astronomy Data Analysis (CIRADA). CIRADA is funded by a grant from the Canada Foundation for Innovation 2017 Innovation Fund (Project 35999), as well as by the Provinces of Ontario, British Columbia, Alberta, Manitoba and Quebec, in collaboration with the National Research Council of Canada, the US National Radio Astronomy Observatory and Australia’s Commonwealth Scientific and Industrial Research Organisation. 
The processing of LOFAR data was enabled by resources provided by the Swedish National Infrastructure for Computing (SNIC) at Chalmers Centre for Computational Science and Engineering (C3SE) partially funded by the Swedish Research Council through grant agreement no.~2018-05973. 
SPO acknowledges support from the Comunidad de Madrid Atracción de Talento program via grant 2022-T1/TIC-23797. 
VV acknowledges support from the Premio per Giovani Ricercatori “Gianni Tofani” II edizione, promoted by INAF-Osservatorio Astrofisico di Arcetri (DD n. 84/2023).
IP acknowledges support from INAF under the Large Grant 2022 funding scheme (project "MeerKAT and LOFAR Team up: a Unique Radio Window on Galaxy/AGN co-Evolution”).    
\end{acknowledgements}

\bibliographystyle{aa}
\bibliography{aanda.bib}

\begin{thebibliography}{61}
\expandafter\ifx\csname natexlab\endcsname\relax\def\natexlab#1{#1}\fi

\bibitem[{{Ahn} {et~al.}(2012){Ahn}, {Alexandroff}, {Allende Prieto}, {Anderson}, {Anderton}, {Andrews}, {Aubourg}, {Bailey}, {Balbinot}, {Barnes}, {Bautista}, {Beers}, {Beifiori}, {Berlind}, {Bhardwaj}, {Bizyaev}, {Blake}, {Blanton}, {Blomqvist}, {Bochanski}, {Bolton}, {Borde}, {Bovy}, {Brandt}, {Brinkmann}, {Brown}, {Brownstein}, {Bundy}, {Busca}, {Carithers}, {Carnero}, {Carr}, {Casetti-Dinescu}, {Chen}, {Chiappini}, {Comparat}, {Connolly}, {Crepp}, {Cristiani}, {Croft}, {Cuesta}, {da Costa}, {Davenport}, {Dawson}, {de Putter}, {De Lee}, {Delubac}, {Dhital}, {Ealet}, {Ebelke}, {Edmondson}, {Eisenstein}, {Escoffier}, {Esposito}, {Evans}, {Fan}, {Femen{\'\i}a Castell{\'a}}, {Fern{\'a}ndez Alvar}, {Ferreira}, {Filiz Ak}, {Finley}, {Fleming}, {Font-Ribera}, {Frinchaboy}, {Garc{\'\i}a-Hern{\'a}ndez}, {Garc{\'\i}a P{\'e}rez}, {Ge}, {G{\'e}nova-Santos}, {Gillespie}, {Girardi}, {Gonz{\'a}lez Hern{\'a}ndez}, {Grebel}, {Gunn}, {Guo}, {Haggard}, {Hamilton}, {Harris}, {Hawley}, {Hearty}, {Ho}, {Hogg}, {Holtzman},
  {Honscheid}, {Huehnerhoff}, {Ivans}, {Ivezi{\'c}}, {Jacobson}, {Jiang}, {Johansson}, {Johnson}, {Kauffmann}, {Kirkby}, {Kirkpatrick}, {Klaene}, {Knapp}, {Kneib}, {Le Goff}, {Leauthaud}, {Lee}, {Lee}, {Long}, {Loomis}, {Lucatello}, {Lundgren}, {Lupton}, {Ma}, {Ma}, {MacDonald}, {Mack}, {Mahadevan}, {Maia}, {Majewski}, {Makler}, {Malanushenko}, {Malanushenko}, {Manchado}, {Mandelbaum}, {Manera}, {Maraston}, {Margala}, {Martell}, {McBride}, {McGreer}, {McMahon}, {M{\'e}nard}, {Meszaros}, {Miralda-Escud{\'e}}, {Montero-Dorta}, {Montesano}, {Morrison}, {Muna}, {Munn}, {Murayama}, {Myers}, {Neto}, {Nguyen}, {Nichol}, {Nidever}, {Noterdaeme}, {Nuza}, {Ogando}, {Olmstead}, {Oravetz}, {Owen}, {Padmanabhan}, {Palanque-Delabrouille}, {Pan}, {Parejko}, {Parihar}, {P{\^a}ris}, {Pattarakijwanich}, {Pepper}, {Percival}, {P{\'e}rez-Fournon}, {P{\'e}rez-R{\`a}fols}, {Petitjean}, {Pforr}, {Pieri}, {Pinsonneault}, {Porto de Mello}, {Prada}, {Price-Whelan}, {Raddick}, {Rebolo}, {Rich}, {Richards}, {Robin}, {Rocha-Pinto},
  {Rockosi}, {Roe}, {Ross}, {Ross}, {Rossi}, {Rubi{\~n}o-Martin}, {Samushia}, {Sanchez Almeida}, {S{\'a}nchez}, {Santiago}, {Sayres}, {Schlegel}, {Schlesinger}, {Schmidt}, {Schneider}, {Schultheis}, {Schwope}, {Sc{\'o}ccola}, {Seljak}, {Sheldon}, {Shen}, {Shu}, {Simmerer}, {Simmons}, {Skibba}, {Skrutskie}, {Slosar}, {Sobreira}, {Sobeck}, {Stassun}, {Steele}, {Steinmetz}, {Strauss}, {Streblyanska}, {Suzuki}, {Swanson}, {Tal}, {Thakar}, {Thomas}, {Thompson}, {Tinker}, {Tojeiro}, {Tremonti}, {Vargas Maga{\~n}a}, {Verde}, {Viel}, {Vikas}, {Vogt}, {Wake}, {Wang}, {Weaver}, {Weinberg}, {Weiner}, {West}, {White}, {Wilson}, {Wisniewski}, {Wood-Vasey}, {Yanny}, {Y{\`e}che}, {York}, {Zamora}, {Zasowski}, {Zehavi}, {Zhao}, {Zheng}, {Zhu}, \& {Zinn}}]{Ahn2012}
{Ahn}, C.~P., {Alexandroff}, R., {Allende Prieto}, C., {et~al.} 2012, \apjs, 203, 21

\bibitem[{{Ahumada} {et~al.}(2020){Ahumada}, {Allende Prieto}, {Almeida}, {Anders}, {Anderson}, {Andrews}, {Anguiano}, {Arcodia}, {Armengaud}, {Aubert}, {Avila}, {Avila-Reese}, {Badenes}, {Balland}, {Barger}, {Barrera-Ballesteros}, {Basu}, {Bautista}, {Beaton}, {Beers}, {Benavides}, {Bender}, {Bernardi}, {Bershady}, {Beutler}, {Bidin}, {Bird}, {Bizyaev}, {Blanc}, {Blanton}, {Boquien}, {Borissova}, {Bovy}, {Brandt}, {Brinkmann}, {Brownstein}, {Bundy}, {Bureau}, {Burgasser}, {Burtin}, {Cano-D{\'\i}az}, {Capasso}, {Cappellari}, {Carrera}, {Chabanier}, {Chaplin}, {Chapman}, {Cherinka}, {Chiappini}, {Doohyun Choi}, {Chojnowski}, {Chung}, {Clerc}, {Coffey}, {Comerford}, {Comparat}, {da Costa}, {Cousinou}, {Covey}, {Crane}, {Cunha}, {Ilha}, {Dai}, {Damsted}, {Darling}, {Davidson}, {Davies}, {Dawson}, {De}, {de la Macorra}, {De Lee}, {Queiroz}, {Deconto Machado}, {de la Torre}, {Dell'Agli}, {du Mas des Bourboux}, {Diamond-Stanic}, {Dillon}, {Donor}, {Drory}, {Duckworth}, {Dwelly}, {Ebelke}, {Eftekharzadeh}, {Davis
  Eigenbrot}, {Elsworth}, {Eracleous}, {Erfanianfar}, {Escoffier}, {Fan}, {Farr}, {Fern{\'a}ndez-Trincado}, {Feuillet}, {Finoguenov}, {Fofie}, {Fraser-McKelvie}, {Frinchaboy}, {Fromenteau}, {Fu}, {Galbany}, {Garcia}, {Garc{\'\i}a-Hern{\'a}ndez}, {Garma Oehmichen}, {Ge}, {Geimba Maia}, {Geisler}, {Gelfand}, {Goddy}, {Gonzalez-Perez}, {Grabowski}, {Green}, {Grier}, {Guo}, {Guy}, {Harding}, {Hasselquist}, {Hawken}, {Hayes}, {Hearty}, {Hekker}, {Hogg}, {Holtzman}, {Horta}, {Hou}, {Hsieh}, {Huber}, {Hunt}, {Ider Chitham}, {Imig}, {Jaber}, {Jimenez Angel}, {Johnson}, {Jones}, {J{\"o}nsson}, {Jullo}, {Kim}, {Kinemuchi}, {Kirkpatrick}, {Kite}, {Klaene}, {Kneib}, {Kollmeier}, {Kong}, {Kounkel}, {Krishnarao}, {Lacerna}, {Lan}, {Lane}, {Law}, {Le Goff}, {Leung}, {Lewis}, {Li}, {Lian}, {Lin}, {Long}, {Longa-Pe{\~n}a}, {Lundgren}, {Lyke}, {Mackereth}, {MacLeod}, {Majewski}, {Manchado}, {Maraston}, {Martini}, {Masseron}, {Masters}, {Mathur}, {McDermid}, {Merloni}, {Merrifield}, {M{\'e}sz{\'a}ros}, {Miglio}, {Minniti},
  {Minsley}, {Miyaji}, {Mohammad}, {Mosser}, {Mueller}, {Muna}, {Mu{\~n}oz-Guti{\'e}rrez}, {Myers}, {Nadathur}, {Nair}, {Nandra}, {Correa do Nascimento}, {Nevin}, {Newman}, {Nidever}, {Nitschelm}, {Noterdaeme}, {O'Connell}, {Olmstead}, {Oravetz}, {Oravetz}, {Osorio}, {Pace}, {Padilla}, {Palanque-Delabrouille}, {Palicio}, {Pan}, {Pan}, {Parker}, {Paviot}, {Peirani}, {Ram{\'r}ez}, {Penny}, {Percival}, {Perez-Fournon}, {P{\'e}rez-R{\`a}fols}, {Petitjean}, {Pieri}, {Pinsonneault}, {Poovelil}, {Povick}, {Prakash}, {Price-Whelan}, {Raddick}, {Raichoor}, {Ray}, {Rembold}, {Rezaie}, {Riffel}, {Riffel}, {Rix}, {Robin}, {Roman-Lopes}, {Rom{\'a}n-Z{\'u}{\~n}iga}, {Rose}, {Ross}, {Rossi}, {Rowlands}, {Rubin}, {Salvato}, {S{\'a}nchez}, {S{\'a}nchez-Menguiano}, {S{\'a}nchez-Gallego}, {Sayres}, {Schaefer}, {Schiavon}, {Schimoia}, {Schlafly}, {Schlegel}, {Schneider}, {Schultheis}, {Schwope}, {Seo}, {Serenelli}, {Shafieloo}, {Shamsi}, {Shao}, {Shen}, {Shetrone}, {Shirley}, {Silva Aguirre}, {Simon}, {Skrutskie}, {Slosar},
  {Smethurst}, {Sobeck}, {Sodi}, {Souto}, {Stark}, {Stassun}, {Steinmetz}, {Stello}, {Stermer}, {Storchi-Bergmann}, {Streblyanska}, {Stringfellow}, {Stutz}, {Su{\'a}rez}, {Sun}, {Taghizadeh-Popp}, {Talbot}, {Tayar}, {Thakar}, {Theriault}, {Thomas}, {Thomas}, {Tinker}, {Tojeiro}, {Toledo}, {Tremonti}, {Troup}, {Tuttle}, {Unda-Sanzana}, {Valentini}, {Vargas-Gonz{\'a}lez}, {Vargas-Maga{\~n}a}, {V{\'a}zquez-Mata}, {Vivek}, {Wake}, {Wang}, {Weaver}, {Weijmans}, {Wild}, {Wilson}, {Wilson}, {Wolthuis}, {Wood-Vasey}, {Yan}, {Yang}, {Y{\`e}che}, {Zamora}, {Zarrouk}, {Zasowski}, {Zhang}, {Zhao}, {Zhao}, {Zheng}, {Zheng}, {Zhu}, \& {Zou}}]{Ahumada2020ApJS..249....3A}
{Ahumada}, R., {Allende Prieto}, C., {Almeida}, A., {et~al.} 2020, \apjs, 249, 3

\bibitem[{{An} {et~al.}(2024){An}, {Vaccari}, {Best}, {Ocran}, {Ishwara-Chandra}, {Taylor}, {Leslie}, {R{\"o}ttgering}, {Kondapally}, {Haskell}, {Collier}, \& {Bonato}}]{An2024MNRAS.528.5346A}
{An}, F., {Vaccari}, M., {Best}, P.~N., {et~al.} 2024, \mnras, 528, 5346

\bibitem[{{Astropy Collaboration} {et~al.}(2013){Astropy Collaboration}, {Robitaille}, {Tollerud}, {Greenfield}, {Droettboom}, {Bray}, {Aldcroft}, {Davis}, {Ginsburg}, {Price-Whelan}, {Kerzendorf}, {Conley}, {Crighton}, {Barbary}, {Muna}, {Ferguson}, {Grollier}, {Parikh}, {Nair}, {Unther}, {Deil}, {Woillez}, {Conseil}, {Kramer}, {Turner}, {Singer}, {Fox}, {Weaver}, {Zabalza}, {Edwards}, {Azalee Bostroem}, {Burke}, {Casey}, {Crawford}, {Dencheva}, {Ely}, {Jenness}, {Labrie}, {Lim}, {Pierfederici}, {Pontzen}, {Ptak}, {Refsdal}, {Servillat}, \& {Streicher}}]{astropy}
{Astropy Collaboration}, {Robitaille}, T.~P., {Tollerud}, E.~J., {et~al.} 2013, \aap, 558, A33

\bibitem[{{Berger} {et~al.}(2021){Berger}, {Adebahr}, {Herrera Ruiz}, {Wright}, {Prandoni}, \& {Dettmar}}]{Berger2021}
{Berger}, A., {Adebahr}, B., {Herrera Ruiz}, N., {et~al.} 2021, \aap, 653, A155

\bibitem[{{B{\"o}hringer} {et~al.}(2000){B{\"o}hringer}, {Voges}, {Huchra}, {McLean}, {Giacconi}, {Rosati}, {Burg}, {Mader}, {Schuecker}, {Simi{\c{c}}}, {Komossa}, {Reiprich}, {Retzlaff}, \& {Tr{\"u}mper}}]{Boringer2000ApJS..129..435B}
{B{\"o}hringer}, H., {Voges}, W., {Huchra}, J.~P., {et~al.} 2000, \apjs, 129, 435

\bibitem[{{Bonafede} {et~al.}(2011){Bonafede}, {Govoni}, {Feretti}, {Murgia}, {Giovannini}, \& {Br{\"u}ggen}}]{Bonafede2011}
{Bonafede}, A., {Govoni}, F., {Feretti}, L., {et~al.} 2011, \aap, 530, A24

\bibitem[{{Carretti} {et~al.}(2023){Carretti}, {O'Sullivan}, {Vacca}, {Vazza}, {Gheller}, {Vernstrom}, \& {Bonafede}}]{Carretti2023}
{Carretti}, E., {O'Sullivan}, S.~P., {Vacca}, V., {et~al.} 2023, \mnras, 518, 2273

\bibitem[{{Carretti} {et~al.}(2022){Carretti}, {Vacca}, {O'Sullivan}, {Heald}, {Horellou}, {R{\"o}ttgering}, {Scaife}, {Shimwell}, {Shulevski}, {Stuardi}, \& {Vernstrom}}]{Carretti2022}
{Carretti}, E., {Vacca}, V., {O'Sullivan}, S.~P., {et~al.} 2022, \mnras, 512, 945

\bibitem[{{Chow-Mart{\'\i}nez} {et~al.}(2014){Chow-Mart{\'\i}nez}, {Andernach}, {Caretta}, \& {Trejo-Alonso}}]{ChowMartinez2014}
{Chow-Mart{\'\i}nez}, M., {Andernach}, H., {Caretta}, C.~A., \& {Trejo-Alonso}, J.~J. 2014, \mnras, 445, 4073

\bibitem[{{Clarke}(2004)}]{Clarke2004}
{Clarke}, T.~E. 2004, Journal of Korean Astronomical Society, 37, 337

\bibitem[{{Condon} {et~al.}(1998){Condon}, {Cotton}, {Greisen}, {Yin}, {Perley}, {Taylor}, \& {Broderick}}]{condon1998}
{Condon}, J.~J., {Cotton}, W.~D., {Greisen}, E.~W., {et~al.} 1998, \aj, 115, 1693

\bibitem[{{Condon} \& {Matthews}(2018)}]{Condon2018}
{Condon}, J.~J. \& {Matthews}, A.~M. 2018, \pasp, 130, 073001

\bibitem[{{DESI Collaboration} {et~al.}(2024){DESI Collaboration}, {Adame}, {Aguilar}, {Ahlen}, {Alam}, {Aldering}, {Alexander}, {Alfarsy}, {Prieto}, {Alvarez}, {Alves}, {Anand}, {Andrade-Oliveira}, {Armengaud}, {Asorey}, {Avila}, {Aviles}, {Bailey}, {Balaguera-Antol{\'\i}nez}, {Ballester}, {Baltay}, {Bault}, {Bautista}, {Behera}, {Beltran}, {BenZvi}, {Beraldo e Silva}, {Bermejo-Climent}, {Berti}, {Besuner}, {Beutler}, {Bianchi}, {Blake}, {Blum}, {Bolton}, {Brieden}, {Brodzeller}, {Brooks}, {Brown}, {Buckley-Geer}, {Burtin}, {Cabayol-Garcia}, {Cai}, {Canning}, {Cardiel-Sas}, {Carnero Rosell}, {Castander}, {Cervantes-Cota}, {Chabanier}, {Chaussidon}, {Chaves-Montero}, {Chen}, {Chen}, {Chuang}, {Claybaugh}, {Cole}, {Cooper}, {Cuceu}, {Davis}, {Dawson}, {de Belsunce}, {de la Cruz}, {de la Macorra}, {Della Costa}, {de Mattia}, {Demina}, {Demirbozan}, {DeRose}, {Dey}, {Dey}, {Dhungana}, {Ding}, {Ding}, {Doel}, {Doshi}, {Douglass}, {Edge}, {Eftekharzadeh}, {Eisenstein}, {Elliott}, {Ereza}, {Escoffier}, {Fagrelius},
  {Fan}, {Fanning}, {Fawcett}, {Ferraro}, {Flaugher}, {Font-Ribera}, {Forero-Romero}, {Forero-S{\'a}nchez}, {Frenk}, {G{\"a}nsicke}, {Garc{\'\i}a}, {Garc{\'\i}a-Bellido}, {Garcia-Quintero}, {Garrison}, {Gil-Mar{\'\i}n}, {Golden-Marx}, {Gontcho}, {Gonzalez-Morales}, {Gonzalez-Perez}, {Gordon}, {Graur}, {Green}, {Gruen}, {Guy}, {Hadzhiyska}, {Hahn}, {Han}, {Hanif}, {Herrera-Alcantar}, {Honscheid}, {Hou}, {Howlett}, {Huterer}, {Ir{\v{s}}i{\v{c}}}, {Ishak}, {Jacques}, {Jana}, {Jiang}, {Jimenez}, {Jing}, {Joudaki}, {Joyce}, {Jullo}, {Juneau}, {Kara{\c{c}}ayl{\i}}, {Karim}, {Kehoe}, {Kent}, {Khederlarian}, {Kim}, {Kirkby}, {Kisner}, {Kitaura}, {Kizhuprakkat}, {Kneib}, {Koposov}, {Kov{\'a}cs}, {Kremin}, {Krolewski}, {L'Huillier}, {Lahav}, {Lambert}, {Lamman}, {Lan}, {Landriau}, {Lang}, {Lange}, {Lasker}, {Leauthaud}, {Le Guillou}, {Levi}, {Li}, {Linder}, {Lyons}, {Magneville}, {Manera}, {Manser}, {Margala}, {Martini}, {McDonald}, {Medina}, {Medina-Varela}, {Meisner}, {Mena-Fern{\'a}ndez}, {Meneses-Rizo}, {Mezcua},
  {Miquel}, {Montero-Camacho}, {Moon}, {Moore}, {Moustakas}, {Mueller}, {Mundet}, {Mu{\~n}oz-Guti{\'e}rrez}, {Myers}, {Nadathur}, {Napolitano}, {Neveux}, {Newman}, {Nie}, {Nikutta}, {Niz}, {Norberg}, {Noriega}, {Paillas}, {Palanque-Delabrouille}, {Palmese}, {Pan}, {Parkinson}, {Penmetsa}, {Percival}, {P{\'e}rez-Fern{\'a}ndez}, {P{\'e}rez-R{\`a}fols}, {Pieri}, {Poppett}, {Porredon}, {Pothier}, {Prada}, {Pucha}, {Raichoor}, {Ram{\'\i}rez-P{\'e}rez}, {Ramirez-Solano}, {Rashkovetskyi}, {Ravoux}, {Rocher}, {Rockosi}, {Ross}, {Rossi}, {Ruggeri}, {Ruhlmann-Kleider}, {Sabiu}, {Said}, {Saintonge}, {Samushia}, {Sanchez}, {Saulder}, {Schaan}, {Schlafly}, {Schlegel}, {Scholte}, {Schubnell}, {Seo}, {Shafieloo}, {Sharples}, {Sheu}, {Silber}, {Sinigaglia}, {Siudek}, {Slepian}, {Smith}, {Soumagnac}, {Sprayberry}, {Stephey}, {Su{\'a}rez-P{\'e}rez}, {Sun}, {Tan}, {Tarl{\'e}}, {Tojeiro}, {Ure{\~n}a-L{\'o}pez}, {Vaisakh}, {Valcin}, {Valdes}, {Valluri}, {Vargas-Maga{\~n}a}, {Variu}, {Verde}, {Walther}, {Wang}, {Wang}, {Weaver},
  {Weaverdyck}, {Wechsler}, {White}, {Xie}, {Yang}, {Y{\`e}che}, {Yu}, {Yuan}, {Zhang}, {Zhang}, {Zhao}, {Zheng}, {Zhou}, {Zhou}, {Zou}, {Zou}, \& {Zu}}]{DESI2024AJ....168...58D}
{DESI Collaboration}, {Adame}, A.~G., {Aguilar}, J., {et~al.} 2024, \aj, 168, 58

\bibitem[{{Duncan} {et~al.}(2021){Duncan}, {Kondapally}, {Brown}, {Bonato}, {Best}, {R{\"o}ttgering}, {Bondi}, {Bowler}, {Cochrane}, {G{\"u}rkan}, {Hardcastle}, {Jarvis}, {Kunert-Bajraszewska}, {Leslie}, {Ma{\l}ek}, {Morabito}, {O'Sullivan}, {Prandoni}, {Sabater}, {Shimwell}, {Smith}, {Wang}, {Wo{\l}owska}, \& {Tasse}}]{Duncan2021}
{Duncan}, K.~J., {Kondapally}, R., {Brown}, M.~J.~I., {et~al.} 2021, \aap, 648, A4

\bibitem[{{Einasto} {et~al.}(2001){Einasto}, {Einasto}, {Tago}, {M{\"u}ller}, \& {Andernach}}]{Einasto2001AJ....122.2222E}
{Einasto}, M., {Einasto}, J., {Tago}, E., {M{\"u}ller}, V., \& {Andernach}, H. 2001, \aj, 122, 2222

\bibitem[{{Ettori} \& {Balestra}(2009)}]{Ettori2009A&A...496..343E}
{Ettori}, S. \& {Balestra}, I. 2009, \aap, 496, 343

\bibitem[{{Eyles} {et~al.}(2020){Eyles}, {Birkinshaw}, {Smol{\v{c}}i{\'c}}, {Horellou}, {Huynh}, {Butler}, {Delhaize}, {Vignali}, \& {Pierre}}]{Eyles2020}
{Eyles}, R.~A.~J., {Birkinshaw}, M., {Smol{\v{c}}i{\'c}}, V., {et~al.} 2020, \aap, 633, A6

\bibitem[{{Falco} {et~al.}(1998){Falco}, {Kochanek}, \& {Mu{\~n}oz}}]{Falco1998}
{Falco}, E.~E., {Kochanek}, C.~S., \& {Mu{\~n}oz}, J.~A. 1998, \apj, 494, 47

\bibitem[{{Fanaroff} \& {Riley}(1974)}]{FanaroffRiley1974MNRAS.167P..31F}
{Fanaroff}, B.~L. \& {Riley}, J.~M. 1974, \mnras, 167, 31P

\bibitem[{{Govoni} \& {Feretti}(2004)}]{Govoni2004}
{Govoni}, F. \& {Feretti}, L. 2004, International Journal of Modern Physics D, 13, 1549

\bibitem[{{Grant} {et~al.}(2010){Grant}, {Taylor}, {Stil}, {Landecker}, {Kothes}, {Ransom}, \& {Scott}}]{grant2010}
{Grant}, J.~K., {Taylor}, A.~R., {Stil}, J.~M., {et~al.} 2010, \apj, 714, 1689

\bibitem[{{Hammond} {et~al.}(2012){Hammond}, {Robishaw}, \& {Gaensler}}]{Hammond2012arXiv1209.1438H}
{Hammond}, A.~M., {Robishaw}, T., \& {Gaensler}, B.~M. 2012, arXiv e-prints, arXiv:1209.1438

\bibitem[{{Herrera Ruiz} {et~al.}(2021){Herrera Ruiz}, {O'Sullivan}, {Vacca}, {Jeli{\'c}}, {Nikiel-Wroczy{\'n}ski}, {Bourke}, {Sabater}, {Dettmar}, {Heald}, {Horellou}, {Piras}, {Sobey}, {Shimwell}, {Tasse}, {Hardcastle}, {Kondapally}, {Chy{\.z}y}, {Iacobelli}, {Best}, {Br{\"u}ggen}, {Carretti}, \& {Prandoni}}]{HerreraRuiz2021}
{Herrera Ruiz}, N., {O'Sullivan}, S.~P., {Vacca}, V., {et~al.} 2021, \aap, 648, A12

\bibitem[{{Hunter}(2007)}]{Hunter2007}
{Hunter}, J.~D. 2007, Computing in Science and Engineering, 9, 90

\bibitem[{{Hutschenreuter} {et~al.}(2022){Hutschenreuter}, {Anderson}, {Betti}, {Bower}, {Brown}, {Br{\"u}ggen}, {Carretti}, {Clarke}, {Clegg}, {Costa}, {Croft}, {Van Eck}, {Gaensler}, {de Gasperin}, {Haverkorn}, {Heald}, {Hull}, {Inoue}, {Johnston-Hollitt}, {Kaczmarek}, {Law}, {Ma}, {MacMahon}, {Mao}, {Riseley}, {Roy}, {Shanahan}, {Shimwell}, {Stil}, {Sobey}, {O'Sullivan}, {Tasse}, {Vacca}, {Vernstrom}, {Williams}, {Wright}, \& {En{\ss}lin}}]{Hutschenreuter2022}
{Hutschenreuter}, S., {Anderson}, C.~S., {Betti}, S., {et~al.} 2022, \aap, 657, A43

\bibitem[{{Kondapally} {et~al.}(2021){Kondapally}, {Best}, {Hardcastle}, {Nisbet}, {Bonato}, {Sabater}, {Duncan}, {McCheyne}, {Cochrane}, {Bowler}, {Williams}, {Shimwell}, {Tasse}, {Croston}, {Goyal}, {Jamrozy}, {Jarvis}, {Mahatma}, {R{\"o}ttgering}, {Smith}, {Wo{\l}owska}, {Bondi}, {Brienza}, {Brown}, {Br{\"u}ggen}, {Chambers}, {Garrett}, {G{\"u}rkan}, {Huber}, {Kunert-Bajraszewska}, {Magnier}, {Mingo}, {Mostert}, {Nikiel-Wroczy{\'n}ski}, {O'Sullivan}, {Paladino}, {Ploeckinger}, {Prandoni}, {Rosenthal}, {Schwarz}, {Shulevski}, {Wagenveld}, \& {Wang}}]{kondapally2021}
{Kondapally}, R., {Best}, P.~N., {Hardcastle}, M.~J., {et~al.} 2021, \aap, 648, A3

\bibitem[{{Lacy} {et~al.}(2020){Lacy}, {Baum}, {Chandler}, {Chatterjee}, {Clarke}, {Deustua}, {English}, {Farnes}, {Gaensler}, {Gugliucci}, {Hallinan}, {Kent}, {Kimball}, {Law}, {Lazio}, {Marvil}, {Mao}, {Medlin}, {Mooley}, {Murphy}, {Myers}, {Osten}, {Richards}, {Rosolowsky}, {Rudnick}, {Schinzel}, {Sivakoff}, {Sjouwerman}, {Taylor}, {White}, {Wrobel}, {Andernach}, {Beasley}, {Berger}, {Bhatnager}, {Birkinshaw}, {Bower}, {Brandt}, {Brown}, {Burke-Spolaor}, {Butler}, {Comerford}, {Demorest}, {Fu}, {Giacintucci}, {Golap}, {G{\"u}th}, {Hales}, {Hiriart}, {Hodge}, {Horesh}, {Ivezi{\'c}}, {Jarvis}, {Kamble}, {Kassim}, {Liu}, {Loinard}, {Lyons}, {Masters}, {Mezcua}, {Moellenbrock}, {Mroczkowski}, {Nyland}, {O'Dea}, {O'Sullivan}, {Peters}, {Radford}, {Rao}, {Robnett}, {Salcido}, {Shen}, {Sobotka}, {Witz}, {Vaccari}, {van Weeren}, {Vargas}, {Williams}, \& {Yoon}}]{Lacy2020}
{Lacy}, M., {Baum}, S.~A., {Chandler}, C.~J., {et~al.} 2020, \pasp, 132, 035001

\bibitem[{{Mahatma} {et~al.}(2021){Mahatma}, {Hardcastle}, {Harwood}, {O'Sullivan}, {Heald}, {Horellou}, \& {Smith}}]{Mahatma2021}
{Mahatma}, V.~H., {Hardcastle}, M.~J., {Harwood}, J., {et~al.} 2021, \mnras, 502, 273

\bibitem[{{Mandal} {et~al.}(2021){Mandal}, {Prandoni}, {Hardcastle}, {Shimwell}, {Intema}, {Tasse}, {van Weeren}, {Algera}, {Emig}, {R{\"o}ttgering}, {Schwarz}, {Siewert}, {Best}, {Bonato}, {Bondi}, {Jarvis}, {Kondapally}, {Leslie}, {Mahatma}, {Sabater}, {Retana-Montenegro}, \& {Williams}}]{Mandal2021}
{Mandal}, S., {Prandoni}, I., {Hardcastle}, M.~J., {et~al.} 2021, \aap, 648, A5

\bibitem[{{Massaro} {et~al.}(2015){Massaro}, {Maselli}, {Leto}, {Marchegiani}, {Perri}, {Giommi}, \& {Piranomonte}}]{Massaro2015Ap&SS.357...75M}
{Massaro}, E., {Maselli}, A., {Leto}, C., {et~al.} 2015, \apss, 357, 75

\bibitem[{{Mingo} {et~al.}(2022){Mingo}, {Croston}, {Best}, {Duncan}, {Hardcastle}, {Kondapally}, {Prandoni}, {Sabater}, {Shimwell}, {Williams}, {Baldi}, {Bonato}, {Bondi}, {Dabhade}, {G{\"u}rkan}, {Ineson}, {Magliocchetti}, {Miley}, {Pierce}, \& {R{\"o}ttgering}}]{Mingo2022}
{Mingo}, B., {Croston}, J.~H., {Best}, P.~N., {et~al.} 2022, \mnras, 511, 3250

\bibitem[{{Mingo} {et~al.}(2019){Mingo}, {Croston}, {Hardcastle}, {Best}, {Duncan}, {Morganti}, {Rottgering}, {Sabater}, {Shimwell}, {Williams}, {Brienza}, {Gurkan}, {Mahatma}, {Morabito}, {Prandoni}, {Bondi}, {Ineson}, \& {Mooney}}]{Mingo2019}
{Mingo}, B., {Croston}, J.~H., {Hardcastle}, M.~J., {et~al.} 2019, \mnras, 488, 2701

\bibitem[{{Osinga} {et~al.}(2022){Osinga}, {van Weeren}, {Andrade-Santos}, {Rudnick}, {Bonafede}, {Clarke}, {Duncan}, {Giacintucci}, {Mroczkowski}, \& {R{\"o}ttgering}}]{Osinga2022A&A...665A..71O}
{Osinga}, E., {van Weeren}, R.~J., {Andrade-Santos}, F., {et~al.} 2022, \aap, 665, A71

\bibitem[{{Osinga} {et~al.}(2021){Osinga}, {van Weeren}, {Boxelaar}, {Brunetti}, {Botteon}, {Br{\"u}ggen}, {Shimwell}, {Bonafede}, {Best}, {Bonato}, {Cassano}, {Gastaldello}, {di Gennaro}, {Hardcastle}, {Mandal}, {Rossetti}, {R{\"o}ttgering}, {Sabater}, \& {Tasse}}]{Osinga2021A&A...648A..11O}
{Osinga}, E., {van Weeren}, R.~J., {Boxelaar}, J.~M., {et~al.} 2021, \aap, 648, A11

\bibitem[{{O'Sullivan} {et~al.}(2020){O'Sullivan}, {Br{\"u}ggen}, {Vazza}, {Carretti}, {Locatelli}, {Stuardi}, {Vacca}, {Vernstrom}, {Heald}, {Horellou}, {Shimwell}, {Hardcastle}, {Tasse}, \& {R{\"o}ttgering}}]{OSullivan2020}
{O'Sullivan}, S.~P., {Br{\"u}ggen}, M., {Vazza}, F., {et~al.} 2020, \mnras, 495, 2607

\bibitem[{{O'Sullivan} {et~al.}(2019){O'Sullivan}, {Machalski}, {Van Eck}, {Heald}, {Br{\"u}ggen}, {Fynbo}, {Heintz}, {Lara-Lopez}, {Vacca}, {Hardcastle}, {Shimwell}, {Tasse}, {Vazza}, {Andernach}, {Birkinshaw}, {Haverkorn}, {Horellou}, {Williams}, {Harwood}, {Brunetti}, {Anderson}, {Mao}, {Nikiel-Wroczy{\'n}ski}, {Takahashi}, {Carretti}, {Vernstrom}, {van Weeren}, {Orr{\'u}}, {Morabito}, \& {Callingham}}]{O'Sullivan2019A&A...622A..16O}
{O'Sullivan}, S.~P., {Machalski}, J., {Van Eck}, C.~L., {et~al.} 2019, \aap, 622, A16

\bibitem[{{O'Sullivan} {et~al.}(2023){O'Sullivan}, {Shimwell}, {Hardcastle}, {Tasse}, {Heald}, {Carretti}, {Br{\"u}ggen}, {Vacca}, {Sobey}, {Van Eck}, {Horellou}, {Beck}, {Bilicki}, {Bourke}, {Botteon}, {Croston}, {Drabent}, {Duncan}, {Heesen}, {Ideguchi}, {Kirwan}, {Lawlor}, {Mingo}, {Nikiel-Wroczy{\'n}ski}, {Piotrowska}, {Scaife}, \& {van Weeren}}]{lotssdr2rm}
{O'Sullivan}, S.~P., {Shimwell}, T.~W., {Hardcastle}, M.~J., {et~al.} 2023, \mnras, 519, 5723

\bibitem[{{Piras} {et~al.}(2024){Piras}, {Horellou}, {Conway}, {Thomasson}, {del Palacio}, {Shimwell}, {O'Sullivan}, {Carretti}, {{\v{S}}nidari{\'c}}, {Jeli{\'c}}, {Adebahr}, {Berger}, {Best}, {Br{\"u}ggen}, {Herrera Ruiz}, {Paladino}, {Prandoni}, {Sabater}, \& {Vacca}}]{piras2024a}
{Piras}, S., {Horellou}, C., {Conway}, J.~E., {et~al.} 2024, \aap, 687, A267

\bibitem[{{Planck Collaboration} {et~al.}(2016{\natexlab{a}}){Planck Collaboration}, {Ade}, {Aghanim}, {Arnaud}, {Ashdown}, {Aumont}, {Baccigalupi}, {Banday}, {Barreiro}, {Barrena}, {Bartlett}, {Bartolo}, {Battaner}, {Battye}, {Benabed}, {Beno{\^\i}t}, {Benoit-L{\'e}vy}, {Bernard}, {Bersanelli}, {Bielewicz}, {Bikmaev}, {B{\"o}hringer}, {Bonaldi}, {Bonavera}, {Bond}, {Borrill}, {Bouchet}, {Bucher}, {Burenin}, {Burigana}, {Butler}, {Calabrese}, {Cardoso}, {Carvalho}, {Catalano}, {Challinor}, {Chamballu}, {Chary}, {Chiang}, {Chon}, {Christensen}, {Clements}, {Colombi}, {Colombo}, {Combet}, {Comis}, {Couchot}, {Coulais}, {Crill}, {Curto}, {Cuttaia}, {Dahle}, {Danese}, {Davies}, {Davis}, {de Bernardis}, {de Rosa}, {de Zotti}, {Delabrouille}, {D{\'e}sert}, {Dickinson}, {Diego}, {Dolag}, {Dole}, {Donzelli}, {Dor{\'e}}, {Douspis}, {Ducout}, {Dupac}, {Efstathiou}, {Eisenhardt}, {Elsner}, {En{\ss}lin}, {Eriksen}, {Falgarone}, {Fergusson}, {Feroz}, {Ferragamo}, {Finelli}, {Forni}, {Frailis}, {Fraisse}, {Franceschi},
  {Frejsel}, {Galeotta}, {Galli}, {Ganga}, {G{\'e}nova-Santos}, {Giard}, {Giraud-H{\'e}raud}, {Gjerl{\o}w}, {Gonz{\'a}lez-Nuevo}, {G{\'o}rski}, {Grainge}, {Gratton}, {Gregorio}, {Gruppuso}, {Gudmundsson}, {Hansen}, {Hanson}, {Harrison}, {Hempel}, {Henrot-Versill{\'e}}, {Hern{\'a}ndez-Monteagudo}, {Herranz}, {Hildebrandt}, {Hivon}, {Hobson}, {Holmes}, {Hornstrup}, {Hovest}, {Huffenberger}, {Hurier}, {Jaffe}, {Jaffe}, {Jin}, {Jones}, {Juvela}, {Keih{\"a}nen}, {Keskitalo}, {Khamitov}, {Kisner}, {Kneissl}, {Knoche}, {Kunz}, {Kurki-Suonio}, {Lagache}, {Lamarre}, {Lasenby}, {Lattanzi}, {Lawrence}, {Leonardi}, {Lesgourgues}, {Levrier}, {Liguori}, {Lilje}, {Linden-V{\o}rnle}, {L{\'o}pez-Caniego}, {Lubin}, {Mac{\'\i}as-P{\'e}rez}, {Maggio}, {Maino}, {Mak}, {Mandolesi}, {Mangilli}, {Martin}, {Mart{\'\i}nez-Gonz{\'a}lez}, {Masi}, {Matarrese}, {Mazzotta}, {McGehee}, {Mei}, {Melchiorri}, {Melin}, {Mendes}, {Mennella}, {Migliaccio}, {Mitra}, {Miville-Desch{\^e}nes}, {Moneti}, {Montier}, {Morgante}, {Mortlock}, {Moss},
  {Munshi}, {Murphy}, {Naselsky}, {Nastasi}, {Nati}, {Natoli}, {Netterfield}, {N{\o}rgaard-Nielsen}, {Noviello}, {Novikov}, {Novikov}, {Olamaie}, {Oxborrow}, {Paci}, {Pagano}, {Pajot}, {Paoletti}, {Pasian}, {Patanchon}, {Pearson}, {Perdereau}, {Perotto}, {Perrott}, {Perrotta}, {Pettorino}, {Piacentini}, {Piat}, {Pierpaoli}, {Pietrobon}, {Plaszczynski}, {Pointecouteau}, {Polenta}, {Pratt}, {Pr{\'e}zeau}, {Prunet}, {Puget}, {Rachen}, {Reach}, {Rebolo}, {Reinecke}, {Remazeilles}, {Renault}, {Renzi}, {Ristorcelli}, {Rocha}, {Rosset}, {Rossetti}, {Roudier}, {Rozo}, {Rubi{\~n}o-Mart{\'\i}n}, {Rumsey}, {Rusholme}, {Rykoff}, {Sandri}, {Santos}, {Saunders}, {Savelainen}, {Savini}, {Schammel}, {Scott}, {Seiffert}, {Shellard}, {Shimwell}, {Spencer}, {Stanford}, {Stern}, {Stolyarov}, {Stompor}, {Streblyanska}, {Sudiwala}, {Sunyaev}, {Sutton}, {Suur-Uski}, {Sygnet}, {Tauber}, {Terenzi}, {Toffolatti}, {Tomasi}, {Tramonte}, {Tristram}, {Tucci}, {Tuovinen}, {Umana}, {Valenziano}, {Valiviita}, {Van Tent}, {Vielva}, {Villa},
  {Wade}, {Wandelt}, {Wehus}, {White}, {Wright}, {Yvon}, {Zacchei}, \& {Zonca}}]{PlanckCollaboration2016XA&A...594A..27P}
{Planck Collaboration}, {Ade}, P.~A.~R., {Aghanim}, N., {et~al.} 2016{\natexlab{a}}, \aap, 594, A27

\bibitem[{{Planck Collaboration} {et~al.}(2016{\natexlab{b}}){Planck Collaboration}, {Ade}, {Aghanim}, {Arnaud}, {Ashdown}, {Aumont}, {Baccigalupi}, {Banday}, {Barreiro}, {Bartlett}, {Bartolo}, {Battaner}, {Battye}, {Benabed}, {Beno{\^\i}t}, {Benoit-L{\'e}vy}, {Bernard}, {Bersanelli}, {Bielewicz}, {Bock}, {Bonaldi}, {Bonavera}, {Bond}, {Borrill}, {Bouchet}, {Boulanger}, {Bucher}, {Burigana}, {Butler}, {Calabrese}, {Cardoso}, {Catalano}, {Challinor}, {Chamballu}, {Chary}, {Chiang}, {Chluba}, {Christensen}, {Church}, {Clements}, {Colombi}, {Colombo}, {Combet}, {Coulais}, {Crill}, {Curto}, {Cuttaia}, {Danese}, {Davies}, {Davis}, {de Bernardis}, {de Rosa}, {de Zotti}, {Delabrouille}, {D{\'e}sert}, {Di Valentino}, {Dickinson}, {Diego}, {Dolag}, {Dole}, {Donzelli}, {Dor{\'e}}, {Douspis}, {Ducout}, {Dunkley}, {Dupac}, {Efstathiou}, {Elsner}, {En{\ss}lin}, {Eriksen}, {Farhang}, {Fergusson}, {Finelli}, {Forni}, {Frailis}, {Fraisse}, {Franceschi}, {Frejsel}, {Galeotta}, {Galli}, {Ganga}, {Gauthier}, {Gerbino}, {Ghosh},
  {Giard}, {Giraud-H{\'e}raud}, {Giusarma}, {Gjerl{\o}w}, {Gonz{\'a}lez-Nuevo}, {G{\'o}rski}, {Gratton}, {Gregorio}, {Gruppuso}, {Gudmundsson}, {Hamann}, {Hansen}, {Hanson}, {Harrison}, {Helou}, {Henrot-Versill{\'e}}, {Hern{\'a}ndez-Monteagudo}, {Herranz}, {Hildebrandt}, {Hivon}, {Hobson}, {Holmes}, {Hornstrup}, {Hovest}, {Huang}, {Huffenberger}, {Hurier}, {Jaffe}, {Jaffe}, {Jones}, {Juvela}, {Keih{\"a}nen}, {Keskitalo}, {Kisner}, {Kneissl}, {Knoche}, {Knox}, {Kunz}, {Kurki-Suonio}, {Lagache}, {L{\"a}hteenm{\"a}ki}, {Lamarre}, {Lasenby}, {Lattanzi}, {Lawrence}, {Leahy}, {Leonardi}, {Lesgourgues}, {Levrier}, {Lewis}, {Liguori}, {Lilje}, {Linden-V{\o}rnle}, {L{\'o}pez-Caniego}, {Lubin}, {Mac{\'\i}as-P{\'e}rez}, {Maggio}, {Maino}, {Mandolesi}, {Mangilli}, {Marchini}, {Maris}, {Martin}, {Martinelli}, {Mart{\'\i}nez-Gonz{\'a}lez}, {Masi}, {Matarrese}, {McGehee}, {Meinhold}, {Melchiorri}, {Melin}, {Mendes}, {Mennella}, {Migliaccio}, {Millea}, {Mitra}, {Miville-Desch{\^e}nes}, {Moneti}, {Montier}, {Morgante},
  {Mortlock}, {Moss}, {Munshi}, {Murphy}, {Naselsky}, {Nati}, {Natoli}, {Netterfield}, {N{\o}rgaard-Nielsen}, {Noviello}, {Novikov}, {Novikov}, {Oxborrow}, {Paci}, {Pagano}, {Pajot}, {Paladini}, {Paoletti}, {Partridge}, {Pasian}, {Patanchon}, {Pearson}, {Perdereau}, {Perotto}, {Perrotta}, {Pettorino}, {Piacentini}, {Piat}, {Pierpaoli}, {Pietrobon}, {Plaszczynski}, {Pointecouteau}, {Polenta}, {Popa}, {Pratt}, {Pr{\'e}zeau}, {Prunet}, {Puget}, {Rachen}, {Reach}, {Rebolo}, {Reinecke}, {Remazeilles}, {Renault}, {Renzi}, {Ristorcelli}, {Rocha}, {Rosset}, {Rossetti}, {Roudier}, {Rouill{\'e} d'Orfeuil}, {Rowan-Robinson}, {Rubi{\~n}o-Mart{\'\i}n}, {Rusholme}, {Said}, {Salvatelli}, {Salvati}, {Sandri}, {Santos}, {Savelainen}, {Savini}, {Scott}, {Seiffert}, {Serra}, {Shellard}, {Spencer}, {Spinelli}, {Stolyarov}, {Stompor}, {Sudiwala}, {Sunyaev}, {Sutton}, {Suur-Uski}, {Sygnet}, {Tauber}, {Terenzi}, {Toffolatti}, {Tomasi}, {Tristram}, {Trombetti}, {Tucci}, {Tuovinen}, {T{\"u}rler}, {Umana}, {Valenziano}, {Valiviita},
  {Van Tent}, {Vielva}, {Villa}, {Wade}, {Wandelt}, {Wehus}, {White}, {White}, {Wilkinson}, {Yvon}, {Zacchei}, \& {Zonca}}]{PlanckCollaboration2016}
{Planck Collaboration}, {Ade}, P.~A.~R., {Aghanim}, N., {et~al.} 2016{\natexlab{b}}, \aap, 594, A13

\bibitem[{{Richards} {et~al.}(2009){Richards}, {Myers}, {Gray}, {Riegel}, {Nichol}, {Brunner}, {Szalay}, {Schneider}, \& {Anderson}}]{Richards2009}
{Richards}, G.~T., {Myers}, A.~D., {Gray}, A.~G., {et~al.} 2009, \apjs, 180, 67

\bibitem[{{Robitaille} \& {Bressert}(2012)}]{RobitailleBressert2012}
{Robitaille}, T. \& {Bressert}, E. 2012, {APLpy: Astronomical Plotting Library in Python}, Astrophysics Source Code Library, record ascl:1208.017

\bibitem[{{Sabater} {et~al.}(2021){Sabater}, {Best}, {Tasse}, {Hardcastle}, {Shimwell}, {Nisbet}, {Jelic}, {Callingham}, {R{\"o}ttgering}, {Bonato}, {Bondi}, {Ciardi}, {Cochrane}, {Jarvis}, {Kondapally}, {Koopmans}, {O'Sullivan}, {Prandoni}, {Schwarz}, {Smith}, {Wang}, {Williams}, \& {Zaroubi}}]{Sabater2021}
{Sabater}, J., {Best}, P.~N., {Tasse}, C., {et~al.} 2021, \aap, 648, A2

\bibitem[{{Sankhyayan} {et~al.}(2023){Sankhyayan}, {Bagchi}, {Tempel}, {More}, {Einasto}, {Dabhade}, {Raychaudhury}, {Athreya}, \& {Hein{\"a}m{\"a}ki}}]{Sankhyayan2023ApJ...958...62S}
{Sankhyayan}, S., {Bagchi}, J., {Tempel}, E., {et~al.} 2023, \apj, 958, 62

\bibitem[{{Shimwell} {et~al.}(2022){Shimwell}, {Hardcastle}, {Tasse}, {Best}, {R{\"o}ttgering}, {Williams}, {Botteon}, {Drabent}, {Mechev}, {Shulevski}, {van Weeren}, {Bester}, {Br{\"u}ggen}, {Brunetti}, {Callingham}, {Chy{\.z}y}, {Conway}, {Dijkema}, {Duncan}, {de Gasperin}, {Hale}, {Haverkorn}, {Hugo}, {Jackson}, {Mevius}, {Miley}, {Morabito}, {Morganti}, {Offringa}, {Oonk}, {Rafferty}, {Sabater}, {Smith}, {Schwarz}, {Smirnov}, {O'Sullivan}, {Vedantham}, {White}, {Albert}, {Alegre}, {Asabere}, {Bacon}, {Bonafede}, {Bonnassieux}, {Brienza}, {Bilicki}, {Bonato}, {Calistro Rivera}, {Cassano}, {Cochrane}, {Croston}, {Cuciti}, {Dallacasa}, {Danezi}, {Dettmar}, {Di Gennaro}, {Edler}, {En{\ss}lin}, {Emig}, {Franzen}, {Garc{\'\i}a-Vergara}, {Grange}, {G{\"u}rkan}, {Hajduk}, {Heald}, {Heesen}, {Hoang}, {Hoeft}, {Horellou}, {Iacobelli}, {Jamrozy}, {Jeli{\'c}}, {Kondapally}, {Kukreti}, {Kunert-Bajraszewska}, {Magliocchetti}, {Mahatma}, {Ma{\l}ek}, {Mandal}, {Massaro}, {Meyer-Zhao}, {Mingo}, {Mostert}, {Nair},
  {Nakoneczny}, {Nikiel-Wroczy{\'n}ski}, {Orr{\'u}}, {Pajdosz-{\'S}mierciak}, {Pasini}, {Prandoni}, {van Piggelen}, {Rajpurohit}, {Retana-Montenegro}, {Riseley}, {Rowlinson}, {Saxena}, {Schrijvers}, {Sweijen}, {Siewert}, {Timmerman}, {Vaccari}, {Vink}, {West}, {Wo{\l}owska}, {Zhang}, \& {Zheng}}]{Shimwell2022}
{Shimwell}, T.~W., {Hardcastle}, M.~J., {Tasse}, C., {et~al.} 2022, \aap, 659, A1

\bibitem[{{Shimwell} {et~al.}(2017){Shimwell}, {R{\"o}ttgering}, {Best}, {Williams}, {Dijkema}, {de Gasperin}, {Hardcastle}, {Heald}, {Hoang}, {Horneffer}, {Intema}, {Mahony}, {Mandal}, {Mechev}, {Morabito}, {Oonk}, {Rafferty}, {Retana-Montenegro}, {Sabater}, {Tasse}, {van Weeren}, {Br{\"u}ggen}, {Brunetti}, {Chy{\.z}y}, {Conway}, {Haverkorn}, {Jackson}, {Jarvis}, {McKean}, {Miley}, {Morganti}, {White}, {Wise}, {van Bemmel}, {Beck}, {Brienza}, {Bonafede}, {Calistro Rivera}, {Cassano}, {Clarke}, {Cseh}, {Deller}, {Drabent}, {van Driel}, {Engels}, {Falcke}, {Ferrari}, {Fr{\"o}hlich}, {Garrett}, {Harwood}, {Heesen}, {Hoeft}, {Horellou}, {Israel}, {Kapi{\'n}ska}, {Kunert-Bajraszewska}, {McKay}, {Mohan}, {Orr{\'u}}, {Pizzo}, {Prandoni}, {Schwarz}, {Shulevski}, {Sipior}, {Smith}, {Sridhar}, {Steinmetz}, {Stroe}, {Varenius}, {van der Werf}, {Zensus}, \& {Zwart}}]{Shimwell2017}
{Shimwell}, T.~W., {R{\"o}ttgering}, H.~J.~A., {Best}, P.~N., {et~al.} 2017, \aap, 598, A104

\bibitem[{{Shimwell} {et~al.}(2019){Shimwell}, {Tasse}, {Hardcastle}, {Mechev}, {Williams}, {Best}, {R{\"o}ttgering}, {Callingham}, {Dijkema}, {de Gasperin}, {Hoang}, {Hugo}, {Mirmont}, {Oonk}, {Prandoni}, {Rafferty}, {Sabater}, {Smirnov}, {van Weeren}, {White}, {Atemkeng}, {Bester}, {Bonnassieux}, {Br{\"u}ggen}, {Brunetti}, {Chy{\.z}y}, {Cochrane}, {Conway}, {Croston}, {Danezi}, {Duncan}, {Haverkorn}, {Heald}, {Iacobelli}, {Intema}, {Jackson}, {Jamrozy}, {Jarvis}, {Lakhoo}, {Mevius}, {Miley}, {Morabito}, {Morganti}, {Nisbet}, {Orr{\'u}}, {Perkins}, {Pizzo}, {Schrijvers}, {Smith}, {Vermeulen}, {Wise}, {Alegre}, {Bacon}, {van Bemmel}, {Beswick}, {Bonafede}, {Botteon}, {Bourke}, {Brienza}, {Calistro Rivera}, {Cassano}, {Clarke}, {Conselice}, {Dettmar}, {Drabent}, {Dumba}, {Emig}, {En{\ss}lin}, {Ferrari}, {Garrett}, {G{\'e}nova-Santos}, {Goyal}, {G{\"u}rkan}, {Hale}, {Harwood}, {Heesen}, {Hoeft}, {Horellou}, {Jackson}, {Kokotanekov}, {Kondapally}, {Kunert-Bajraszewska}, {Mahatma}, {Mahony}, {Mandal}, {McKean},
  {Merloni}, {Mingo}, {Miskolczi}, {Mooney}, {Nikiel-Wroczy{\'n}ski}, {O'Sullivan}, {Quinn}, {Reich}, {Roskowi{\'n}ski}, {Rowlinson}, {Savini}, {Saxena}, {Schwarz}, {Shulevski}, {Sridhar}, {Stacey}, {Urquhart}, {van der Wiel}, {Varenius}, {Webster}, \& {Wilber}}]{Shimwell2019}
{Shimwell}, T.~W., {Tasse}, C., {Hardcastle}, M.~J., {et~al.} 2019, \aap, 622, A1

\bibitem[{{Simonte} {et~al.}(2023){Simonte}, {Andernach}, {Br{\"u}ggen}, {Best}, \& {Osinga}}]{Simonte2023A&A...672A.178S}
{Simonte}, M., {Andernach}, H., {Br{\"u}ggen}, M., {Best}, P.~N., \& {Osinga}, E. 2023, \aap, 672, A178

\bibitem[{{Simonte} {et~al.}(2024){Simonte}, {Andernach}, {Br{\"u}ggen}, {Miley}, \& {Barthel}}]{Simonte2024A&A...686A..21S}
{Simonte}, M., {Andernach}, H., {Br{\"u}ggen}, M., {Miley}, G.~K., \& {Barthel}, P. 2024, \aap, 686, A21

\bibitem[{{Stuardi} {et~al.}(2020){Stuardi}, {O'Sullivan}, {Bonafede}, {Br{\"u}ggen}, {Dabhade}, {Horellou}, {Morganti}, {Carretti}, {Heald}, {Iacobelli}, \& {Vacca}}]{Stuardi2020}
{Stuardi}, C., {O'Sullivan}, S.~P., {Bonafede}, A., {et~al.} 2020, \aap, 638, A48

\bibitem[{{Tarr{\'\i}o} {et~al.}(2019){Tarr{\'\i}o}, {Melin}, \& {Arnaud}}]{Tarrio2019A&A...626A...7T}
{Tarr{\'\i}o}, P., {Melin}, J.~B., \& {Arnaud}, M. 2019, \aap, 626, A7

\bibitem[{{Taylor} {et~al.}(2009){Taylor}, {Stil}, \& {Sunstrum}}]{taylor2009}
{Taylor}, A.~R., {Stil}, J.~M., \& {Sunstrum}, C. 2009, \apj, 702, 1230

\bibitem[{{Taylor}(2005)}]{topcat}
{Taylor}, M.~B. 2005, in Astronomical Society of the Pacific Conference Series, Vol. 347, Astronomical Data Analysis Software and Systems XIV, ed. P.~{Shopbell}, M.~{Britton}, \& R.~{Ebert}, 29

\bibitem[{{Van Eck} {et~al.}(2018){Van Eck}, {Haverkorn}, {Alves}, {Beck}, {Best}, {Carretti}, {Chy{\.z}y}, {Farnes}, {Ferri{\`e}re}, {Hardcastle}, {Heald}, {Horellou}, {Iacobelli}, {Jeli{\'c}}, {Mulcahy}, {O'Sullivan}, {Polderman}, {Reich}, {Riseley}, {R{\"o}ttgering}, {Schnitzeler}, {Shimwell}, {Vacca}, {Vink}, \& {White}}]{VanEck2018}
{Van Eck}, C.~L., {Haverkorn}, M., {Alves}, M.~I.~R., {et~al.} 2018, \aap, 613, A58

\bibitem[{{Vernstrom} {et~al.}(2019){Vernstrom}, {Gaensler}, {Rudnick}, \& {Andernach}}]{Vernstrom2019ApJ...878...92V}
{Vernstrom}, T., {Gaensler}, B.~M., {Rudnick}, L., \& {Andernach}, H. 2019, \apj, 878, 92

\bibitem[{{{\v{S}}nidari{\'c}} {et~al.}(2023){{\v{S}}nidari{\'c}}, {Jeli{\'c}}, {Mevius}, {Brentjens}, {Erceg}, {Shimwell}, {Piras}, {Horellou}, {Sabater}, {Best}, {Bracco}, {Ceraj}, {Haverkorn}, {O'Sullivan}, {Turi{\'c}}, \& {Vacca}}]{Snidaric2023}
{{\v{S}}nidari{\'c}}, I., {Jeli{\'c}}, V., {Mevius}, M., {et~al.} 2023, \aap, 674, A119

\bibitem[{{Wen} \& {Han}(2015)}]{Wen2015ApJ...807..178W}
{Wen}, Z.~L. \& {Han}, J.~L. 2015, \apj, 807, 178

\bibitem[{{Wen} \& {Han}(2024)}]{Wen2024ApJS..272...39W}
{Wen}, Z.~L. \& {Han}, J.~L. 2024, \apjs, 272, 39

\bibitem[{{Yantovski-Barth} {et~al.}(2024){Yantovski-Barth}, {Newman}, {Dey}, {Andrews}, {Eracleous}, {Golden-Marx}, \& {Zhou}}]{Yantovski-Barth2024MNRAS.531.2285Y}
{Yantovski-Barth}, M.~J., {Newman}, J.~A., {Dey}, B., {et~al.} 2024, \mnras, 531, 2285

\bibitem[{{Zou} {et~al.}(2022){Zou}, {Sui}, {Xue}, {Zhou}, {Ma}, {Zhou}, {Nie}, {Zhang}, {Feng}, {Shen}, \& {Wang}}]{Zou2022RAA....22f5001Z}
{Zou}, H., {Sui}, J., {Xue}, S., {et~al.} 2022, Research in Astronomy and Astrophysics, 22, 065001

\end{thebibliography}

\appendix

\twocolumn
\onecolumn
\begin{landscape}
\section{Properties of detected polarized sources}
\begin{table}[ht]   
\label{thebigtable}
\small
\centering 
\caption{Catalog of properties of polarized sources in the ELAIS-N1 LOFAR deep field.}
\label{table:catalog}      
\begin{tabular}{ l r r r r r r c c c c c c}        
\hline\hline  
Source   &  RM        & GRM       &  RRM    & Angular size & Linear size  & Flux    &  log$_{10}$(Luminosity)    &  Redshift       & Redshift        & Host & Radio     \\
   & (rad m$^{-2}$) & (rad m$^{-2}$) & (rad m$^{-2}$) & (arcsec) & (kpc) & (mJy) &  & & type &  classification & morphology  \\
   (1) & (2) &  (3) & (4) & (5) & (6) & (7) & (8) & (9) & (10) & (11) & (12)  \\
\hline
01        &  $1.68 \pm 0.02$  & $1.63 \pm 2.01$  &  $0.05 \pm 2.01$  & 17               & 91              &  163         & 25.89       &  0.37779     & s           & QSO            & FRII      \\    
02$^{(*)}$        &  $9.71 \pm 0.05$  & $4.41 \pm 2.78$  &  $5.30 \pm 2.78$  & 35               & 285             &  471         & 27.30       &  0.98        & p           & QSO            & FRII      \\    
03        &  $-5.83 \pm 0.02$  & $-3.91 \pm 2.69$  &  $-1.92 \pm 2.69$  & 10               & 45              &  602         & 26.23       &  0.3         & s           & Bl             & OE     \\      
04$_{\rm A} $     &  $19.68 \pm 0.01$  & $14.47 \pm 2.83$  &  $5.21 \pm 2.83$  & 77               & 349             &  1737        & 26.66       &  0.29186     & s           & QSO            & FRII    \\      
04$_{\rm B} $      &  $21.78 \pm 0.01$  & $14.47 \pm 2.83$  &  $7.31 \pm 2.83$  & 77               & 349             &  1737        & 26.66       &  0.29186     & s           & QSO            & FRII    \\      
05        &  $19.12 \pm 0.05$  & $9.91 \pm 2.81$  &  $9.21 \pm 2.81$  & 33               & 233             &  196         & 26.49       &  0.6359      & s           & G              & FRII     \\     
06        &  $18.43 \pm 0.03$  & $11.01 \pm 2.84$  &  $7.42 \pm 2.84$  & 300              & 378             &  209         & 24.32       &  0.06333     & s           & G              & FRI      \\     
07        &  $6.062 \pm 0.003$  & $5.17 \pm 2.15$  &  $0.89 \pm 2.15$  & 12               & 91              &  1299        & 27.51       &  0.7683      & s           & G              & FRII        \\  
08        &  $13.39 \pm 0.03$  & $7.66 \pm 2.26$  &  $5.73 \pm 2.26$  & 61               & 521             &  338         & 27.36       &  1.197       & s           & G              & FRII        \\  
09        &  $-3.79 \pm 0.03$  & $2.10 \pm 2.55$  &  $-5.89 \pm 2.55$  & 1                & 9               &  24          & 25.82       &  0.81512     & s           & G              & C          \\   
10        &  $-5.61 \pm 0.05$  & $0.22 \pm 2.43$  &  $-5.83 \pm 2.43$  & 3                & 27              &  122         & 27.40       &  1.94822     & s           & Bl             & C          \\   
11        &  $-6.24 \pm 0.04$  & $0.59 \pm 2.45$  &  $-6.83 \pm 2.45$  & 152              & 1308            &  511         & 27.63       &  1.32046     & s           & QSO            & FRII       \\   
12        &  $7.17 \pm 0.02$  & $7.93 \pm 2.23$  &  $-0.76 \pm 2.23$  & 142              & 1169            &  458         & 27.30       &  0.99258     & s           & QSO            & FRII        \\  
13$_{\rm A}$      &  $2.38 \pm 0.04$  & $5.70 \pm 2.47$  &  $-3.32 \pm 2.47$  & 129              & 956             &  41          & 25.92       &  0.71385     & s           & QSO            & FRII        \\  
14        &  $10.31 \pm 0.02$  & $9.16 \pm 2.93$  &  $1.15 \pm 2.93$  & 53               & 130             &  122         & 24.78       &  0.13455     & s           & G              & FRI         \\  
15$^{(*)}$        &  $-4.80 \pm 0.04$  & $1.49 \pm 1.99$  &  $-6.29 \pm 1.99$  & 96               & 769             &  70          & 26.38       &  0.9         & p           & G              & FRII       \\   
16        &  $-4.01 \pm 0.05$  & $5.17 \pm 2.39$  &  $-9.18 \pm 2.39$  & 45               & 220             &  74          & 25.41       &  0.32936     & s           & G              & OE         \\   
17        &  $1.95 \pm 0.03$  & $2.42 \pm 1.78$  &  $-0.47 \pm 1.78$  & 64               & 315             &  118         & 25.63       &  0.3347      & s           & G              & FRII        \\  
18        &  $-20.31 \pm 0.03$  & $3.78 \pm 2.33$  &  $-24.09 \pm 2.33$  & 144              & 569             &  167         & 25.48       &  0.24297     & s           & G              & FRI       \\    
19        &  $2.91 \pm 0.04$  & $6.46 \pm 2.68$  &  $-3.55 \pm 2.68$  & 30               & 227             &  260         & 26.78       &  0.7523      & s           & G              & FRII     \\     
20        &  $-4.70 \pm 0.02$  & $4.52 \pm 2.34$  &  $-9.22 \pm 2.34$  & 36               & 313             &  144         & 27.34       &  1.6981      & s           & Bl             & OE      \\      
21$^{(*)}$        &  $3.29 \pm 0.04$  & $9.52 \pm 3.19$  &  $-6.23 \pm 3.19$  & 27               & 226             &  348         & 27.28       &  1.09        & p           & G              & FRII      \\    
22        &  $2.31 \pm 0.05$  & $9.79 \pm 3.07$  &  $-7.48 \pm 3.07$  & 15               & 93              &  93          & 25.90       &  0.491       & s           & G              & OE      \\      
23        &  $5.91 \pm 0.05$  & $6.15 \pm 2.63$  &  $-0.24 \pm 2.63$  & 257              & 2089            &  67          & 26.41       &  0.9443      & s           & G              & FRII    \\      
24        &  $9.43 \pm 0.01$  & $8.02 \pm 2.17$  &  $1.41 \pm 2.17$  & 6                & 34              &  258         & 26.18       &  0.415       & p           & Bl             & C      \\       
25        &  $10.03 \pm 0.02$  & $7.46 \pm 2.87$  &  $2.57 \pm 2.87$  & 6                & 53              &  135         & 26.91       &  1.1326      & s           & QSO            & C      \\       
26        &  $-5.38 \pm 0.07$  & $2.72 \pm 2.60$  &  $-8.10 \pm 2.61$  & 158              & 817             &  215         & 25.95       &  0.3566      & s           & QSO            & FRII      \\    
27        &  $14.17 \pm 0.07$  & $9.84 \pm 2.92$  &  $4.33 \pm 2.92$  & 111              & 141             &  190         & 24.30       &  0.06389     & s           & G              & FRII       \\   
28        &  $-5.74 \pm 0.05$  & $2.40 \pm 2.57$  &  $-8.14 \pm 2.57$  & 8                & 51              &  68          & 25.76       &  0.48882     & s           & G              & OE       \\     
29 (13$_{\rm B} $)       &  $3.92 \pm 0.05$  & $5.70 \pm 2.47$  &  $-1.78 \pm 2.47$  & 129              & 956             &  41          & 25.92       &  0.71385     & s           & QSO            & FRII      \\    
30 $^{(*)}$       &  $19.40 \pm 0.07$  & $11.82 \pm 3.45$  &  $7.58 \pm 3.45$  & 30               & 235             &  207         & 26.78       &  0.83        & p           & G              & FRII      \\    
31        &  $6.23 \pm 0.06$  & $8.75 \pm 2.89$  &  $-2.52 \pm 2.89$  & 38               & 315             &  328         & 27.23       &  1.0524      & s           & G              & FRII     \\     
32        &  $-4.05 \pm 0.06$  & $5.74 \pm 2.48$  &  $-9.79 \pm 2.48$  & 140              & 504             &  116         & 25.20       &  0.21431     & s           & G              & FRI   \\
\hline
\end{tabular}
\begin{tablenotes}
\small
\item{
(1) Name of polarized source, as in \citetalias{piras2024a}; 
Source 04$_{\rm A}$ and 04$_{\rm B}$ are two lobes of the same radio galaxy; 
Source 13$_{\rm A}$ and 29 (or 13$_{\rm B}$) are two lobes of the same radio galaxy;
(2) RM and RM error of polarized source, from \citetalias{piras2024a};
(3) GRM and GRM error at the location of the polarized sources;
(4) RRM and RRM error at the location of the polarized sources;
(5) angular size of radio galaxy, including components undetected in polarization;
(6) projected linear size of radio galaxy, including components undetected in polarization;
(7) flux density in Stokes $I$ integrated over the radio source; 
(8) logarithm of rest-frame luminosity in W Hz$^{-1}$ at 146~MHz;
(9) redshift of host galaxy; 
(10) redshift type, p = photometric redshift, s = spectroscopic redshift;
(11) Host classification, QSO = Quasi-Stellar Object, Bl = blazar, G = galaxy;\\
(12) Radio morphology classification at 146~MHz, C = compact, OE = other extended, FRI = FRI radio galaxy, FRII = FRII radio galaxy. \\
$^{(*)}$ The photometric redshifts of these sources are from \cite{Simonte2024A&A...686A..21S}, who determined them by averaging the photometric redshifts from multiple catalogs. \\ 
The full Table and references are available at the CDS. \\
}
\end{tablenotes}
\end{table}
\end{landscape}

\section{Images and Faraday spectra} \label{app:sources_images}
The Stokes $I$ images from \cite{Sabater2021} are shown in contours and the polarized intensity in blue colors. 
The contours increase by factors of $\sqrt{2}$ and 
start at 100 times the noise level of the Stokes~$I$ map, 
$\sigma_I$=30~$\mu$Jy beam$^{-1}$ for sources 02, 03, 04, 05, 07, 08, 09, 10, 19, 24, and 30; 
start at 50$\sigma_I$ for sources 01, 11, 17, 21, 22, 25, 28, and 31; 
start at 20$\sigma_I$ for sources 12, 15, 20, 26, and 32;
start at 5$\sigma_I$ for source~23;
start at 10$\sigma_I$ for the remaining sources. 
The red crosses indicate the locations of the host galaxies. 
The orange square indicate the locations of the detections from \cite{grant2010}. 
The Faraday spectra correspond to the locations of the polarized intensity peaks. 
The instrumental region near 0~rad~m$^{-2}$ is shaded in gray. 
The horizontal red lines show the $8 \sigma_{\rm QU}$ and $6 \sigma_{\rm QU}$ thresholds according to \citetalias{piras2024a}.
In the upper left of each contour map are the source number and the redshift of the source.

\section{Description of the sources} \label{app:sources}
Each detected polarized source was visually inspected in the 6-arcsecond-resolution LOFAR Stokes~$I$ image from \cite{Sabater2021} to assess its morphology. We also inspected Quick Look images from the VLASS taken in different epochs 
(VLASS~1.1, VLASS~2.1, VLASS~3.1; 2-4~GHz band, 2.5~arcsec-resolution; \citealt{Lacy2020}). 
The references for the redshifts quoted below are given in Table~\ref{table:catalog} and are repeated here only in some special cases. 

{Source 01 (ILT J155603.98+550056.8)} is an FRII radio galaxy in the LOFAR image and in VLASS. We detected polarization from the western lobe. 
The host galaxy has been classified as a quasar at spectroscopic redshift $z = 0.37779$.  

{Source 02 (ILT J155614.81+534814.8)} is an FRII radio galaxy both in the LOFAR and in the VLASS image. The polarized emission comes from the southern lobe. The host galaxy has been classified as a quasar at photometric redshift $z = 0.98$. 

{Source 03 (ILT J155848.42+562514.4)} is an extended source in the LOFAR image and is compact in VLASS. 
It is associated with an early-type galaxy with spectroscopic redshift $z = 0.30\pm0.03$ measured from detection of three absorption lines \citep{Falco1998}. This is the redshift value that we used. We note that there are conflicting values in the literature: the SDSS spectrum shows no lines, but a spectroscopic redshift of 3.37 is given in the SDSS DR16 catalog \citep{Ahumada2020ApJS..249....3A}, which might be due to misidentification by the pipeline. The Roma BZCAT multiwavelength catalog of blazars quotes $z = 0$ \citep{Massaro2015Ap&SS.357...75M}.  
It was detected in polarization with LOFAR by \cite{HerreraRuiz2021} and the RM values at 150~MHz are consistent within 1$\sigma$. 

{Source 04 (ILT J160344.42+524228.0)} is an FRII radio galaxy both in the LOFAR and the VLASS images. We detected polarized emission from both lobes with slightly different RM values: 19.68$\pm$0.01~rad~m$^{-2}$ for source 04$_{\rm A}$ and 21.78~$\pm$0.01 rad~m$^{-2}$ for source 04$_{\rm B}$. The RMs agree with the value at 1.4~GHz from the NVSS RM catalog. 
Between the lobes there is a quasar at spectroscopic redshift $z = 0.29186$.

{Source 05 (ILT J160520.16+532837.2)} is FRII radio galaxy both in the LOFAR and the VLASS image. The polarized emission comes from the eastern hotspot. The spectroscopic redshift of the host galaxy is $z = 0.6359$. 

{Source 06 (ILT J160532.84+531257.4)} is an FRI radio galaxy at 150~MHz and a compact source at 3~GHz. The polarized emission comes from the nucleus of the radio galaxy. It was detected by \cite{HerreraRuiz2021} and the RM values at 150~MHz are consistent within 1$\sigma$. 
The associated spectroscopic redshift is $z = 0.06333$. 

{Source 07 (ILT J160538.33+543922.6)}, the reference source that we used in the stacking process (\citetalias{piras2024a}), seems to be an FRII radio galaxy in the LOFAR image, and indeed we see clearly the double-lobed structure in the VLASS image. The polarized emission comes from the south-western component. 
The peculiarity of this source is a star superposed within $1''$ of the peak of SW hotspot in VLASS, which could be mistaken as the host of one-sided source. The real host (which is marked in Fig.~\ref{app:sources_images}) is much fainter than the superposed star.
The RM is consistent with that found by \cite{HerreraRuiz2021}. 
The spectroscopic redshift is $z = 0.7683$. 

{Source 08 (ILT J160603.11+534812.6)} was classified as an FRII radio galaxy at 150~MHz \citep{Mingo2022} and appears as an FRII in the 3~GHz VLASS image as well. We detected polarized emission from the eastern lobe. 
The spectroscopic redshift of the host galaxy is $z = 1.1970$.

{Source 09 (ILT J160725.85+553525.8)} is compact in both the LOFAR and the VLASS images. 
The spectroscopic redshift of the host galaxy is $z = 0.81512$. The Bz

{Source 10 (ILT J160820.72+561355.7)} is compact both at 150~MHz and 3~GHz. 
The host galaxy has been classified as a quasar at spectroscopic redshift $z = 1.9482$ from \cite{Ahumada2020ApJS..249....3A}. It is classified as a blazar in the Roma BZCAT multiwavelength catalog \citep{Massaro2015Ap&SS.357...75M}. 

{Source 11 (ILT J160851.99+561109.3)} was classified as an FRII at LOFAR frequencies \citep{Mingo2022} and resolved as an FRII in the VLASS image, too. We detected polarized emission from the western lobe. 
The host galaxy has been classified as a quasar at spectroscopic redshift $z=1.3204$. 

{Source 12 (ILT J160909.99+535426.8)} is an FRII radio galaxy in the LOFAR image, but the eastern lobe is not visible in the VLASS image. We detected polarized emission from the western lobe. \cite{HerreraRuiz2021} found a RM value of 7.30$\pm$0.02 rad m$^{-2}$, we found 7.17~$\pm$0.02 rad m$^{-2}$. 
The host galaxy has been classified as a quasar at spectroscopic redshift $z=0.99258$ from \cite{Ahumada2020ApJS..249....3A}. It is classified as a blazar in the Roma BZCAT multiwavelength catalog \citep{Massaro2015Ap&SS.357...75M}. 

{Source 13$_{\rm  A}$ (ILT J161118.03+543218.1)} is the northern lobe of an FRII radio galaxy in the LOFAR image \citep{Mingo2022}, but is not detected at 3~GHz. 
The southern lobe is identified as {Source 29} (or {13$_{\rm B}$}) and is polarized in the 6-8$\sigma_{\rm{QU}}$ range. Between the lobes there is a quasar at spectroscopic redshift $z = 0.71385$. 

{Source 14 (ILT J161240.15+533558.3)} is an FRI radio galaxy in the LOFAR image and is a compact source in VLASS. The polarized emission comes from the nucleus of the radio galaxy. The spectroscopic redshift is $z = 0.13455$. 
This source has also been detected by \cite{HerreraRuiz2021} and the RM values are consistent within 2$\sigma$.

{Source 15 (ILT J161314.05+560810.8)} appears as an FRII radio galaxy in the LOFAR image and it is not detected in VLASS. The polarized emission is found on the western lobe. 
Close to the middle between the two lobes the $r=23.37$ magnitude galaxy DESI~J243.2976+56.1347 had been identified by \cite{Simonte2024A&A...686A..21S} as the host, who give $z_{\rm ph} = 0.9\pm0.1$ and a size of 96~arcsec for the entire source. 

{Source 16 (ILT J161530.73+545232.4)} shows diffuse emission at 150~MHz and is not detected at 3~GHz. 
The host galaxy has a spectroscopic redshift $z = 0.32936$. 

{Source 17 (ILT J161548.48+562029.8)} is an FRII radio galaxy in the LOFAR image and in VLASS. 
\cite{HerreraRuiz2021} measured an RM value of 1.43$\pm$0.04 rad~m$^{-2}$, we found 1.95$\pm$0.03 rad m$^{-2}$. Polarized emission is detected in the north-western lobe.
The spectroscopic redshift is $z = 0.3347$. 

{Source 18 (ILT J161623.79+552700.8)} is an FRI radio galaxy with LOFAR \citep{Mingo2022} and it is unresolved in VLASS.  
The polarized emission is associated with the central component of the radio galaxy. It has been detected also by \cite{HerreraRuiz2021} with an RM of $-20.15\pm0.03$ rad~m$^{-2}$; we measured an RM of $-20.31\pm0.03$ rad~m$^{-2}$ in the $6''$ data.
The host galaxy has a spectroscopic redshift $z = 0.24297$. 

{Source 19 (ILT J161832.97+543146.3)} 
is the polarized component associated with the northern hotspot of an FRII radio galaxy. 
The spectroscopic redshift of the galaxy is $z = 0.7523$. 

{Source 20 (ILT J161919.70+553556.7)} was classified as an FRI radio galaxy by \citealt{Mingo2022}. However, this source is very core-dominated in the LOFAR image and seems to have a one-sided jet in the SSW direction, of which just the beginning can be recognized in the VLASS images. 
It is classified as blazar in NED. Calling such a one-sided blazar-like source an FRI may be misleading, so we classified it as ``Other Extended" in Table~\ref{table:catalog}.
The attribution of a redshift is complicated: 
we used $z = 1.698$ (spectroscopic) from \cite{DESI2024AJ....168...58D}, but other values are given in the literature: 
\cite{Duncan2021} gave $z = 0.701$  (photometric);
\cite{HerreraRuiz2021} cited $z = 0.405$ (photometric) from \cite{Richards2009}; 
\cite{Ahn2012} gave $z = 5.5539$ (spectroscopic).  

{Source 21 (ILT J162027.55+534208.8)} is an FRII radio galaxy in the LOFAR and VLASS images. Polarized emission is detected from the south-eastern lobe. This is a $27''$ large asymmetric FRII listed in \citealt{Simonte2024A&A...686A..21S} with $z = 1.09 \pm 0.21$ (photometric) from three references, coincident with a faint VLASS radio core close to the north-western hotspot. 

{Source 22 (ILT J162318.64+533847.4)} is an extended radio source at 150~MHz and 3~GHz. This source is associated with an optical galaxy with spectroscopic redshift $z = 0.4910$.

{Source 23 (ILT J162027.55+534208.8)} is the SW lobe of a $4.3'$ large FRII listed in \cite{Simonte2024A&A...686A..21S} with $z = 0.9443$ (spectroscopic) from SDSS. 
 
{Source 24 (ILT J162432.20+565228.5)} is compact both in LOFAR and VLASS images. It was also detected in polarization by \cite{HerreraRuiz2021} and the RMs found are consistent within 2~$\sigma$. It is cataloged in NED as a blazar and the photometric redshift is 0.415 \citep{Richards2009}. It is classified as a blazar also in the Roma BZCAT multiwavelength catalog \citep{Massaro2015Ap&SS.357...75M}.

{Source 25 (ILT J162634.18+544207.8)} is a compact source in LOFAR and in VLASS. It is identified as a quasar, with spectroscopic redshift of $z = 1.1326$. 

{Source 26 (ILT J160936.45+552659.0)} is an FRII radio galaxy at LOFAR frequencies \citep{Mingo2022}, but is very faint at 3~GHz. 
We detected polarization from the northern lobe.
The host galaxy has been classified as a quasar at spectroscopic redshift $z=0.3566$. 

{Source 27 (ILT J161037.49+532425.1)} appears as an FRII radio galaxy in the LOFAR image, and is barely visible in VLASS. In \cite{Mingo2022} it is classified as an FRI, but it has very diffuse lobes and could also be a remnant FRII. The luminosity we computed ($\sim 2 \times 10^{24}$ W Hz$^{-1}$) is more consistent with an FRI type. However, \cite{Mingo2022} also identified a few FRIIs at similarly low luminosities.
The host galaxy has a spectroscopic redshift of $z = 0.06389$. 

{Source 28 (ILT J161057.72+553527.9)} is an extended source both in the LOFAR image and in VLASS. 
The optical counterpart has a spectroscopic redshift of $z=0.48882$. 

{Source 29, or 13$_{\rm  B}$, (ILT J162347.10+553207.2)} coincides with the southern lobe of an FRII radio galaxy at 150~MHz; it is not visible in the VLASS image at 3~GHz. The polarized emission that we detected from that lobe was in the $6-8 \sigma_{\rm QU}$ range, 
while it was detected above $8 \sigma_{\rm QU}$ in the northern lobe.

{Source 30 (ILT J161340.99+524913.0)} is the north-western lobe of $30''$-wide FRII, very symmetric in VLASS, that had also been identified by \cite{Simonte2024A&A...686A..21S} 
with DESI~J243.4211+52.8203, right at the symmetry center, and with $z=0.83\pm0.05$ (photometric) from four different references. 

{Source 31 (ILT J161537.86+534646.4)} is an FRII radio galaxy at 150~MHz and looks like a FRII at 3~GHz. It is classified as an FRI by \cite{Mingo2022}. 
We detected polarization from the central part of the galaxy, which has a spectroscopic redshift of $z = 1.0524$.

{Source 32 (ILT J161859.41+545246.3)} is an FRI radio galaxy at 150~MHz \citep{Mingo2022}. 
At 3~GHz a central component is clearly visible while a lobe is barely detected. We detected polarized emission from the central component. 
The spectroscopic redshift of the galaxy is $z = 0.21431$. The host galaxy we identified differs from the fainter one (SDSS J161859.62+545246.8) proposed by \cite{Simonte2024A&A...686A..21S}, although both are at the same redshift. However, the wide-angle tailed radio morphology suggests that our identified host is the correct one. It is the brightest cluster galaxy WH J161858.7+545228 listed in \cite{Wen2024ApJS..272...39W}, as reported in Table~\ref{table:clusters}.

\clearpage
\end{document}